\documentclass{aa}  
\usepackage{graphicx}
\usepackage{graphicx,subcaption} 
\usepackage{natbib}
\usepackage{txfonts}

\usepackage{hyperref}

\usepackage[font=small,labelfont=bf]{caption}

\usepackage{newtxtext,newtxmath}
\usepackage[T1]{fontenc}
\usepackage{amsmath}	
\usepackage{booktabs}
\usepackage{multirow}

\titlerunning{An ultraviolet perspective on hierarchical star formation}
\authorrunning{Shashank et al.}

\begin{document} 

\title{ Tracing hierarchical star formation out to kiloparsec scales in nearby spiral galaxies with UVIT }

\author{Gairola Shashank\inst{1,2},
          Smitha Subramanian\inst{1},
          Sreedevi M.\inst{1,3},
          Shyam H 
          Menon\inst{4,5},
          Chayan Mondal\inst{6,7},
          Sriram Krishna\inst{1,2},
          Mousumi Das\inst{1},
          Annapurni Subramaniam\inst{1}}

   \institute{Indian Institute of Astrophysics, Koramangala II Block, Bangalore-560034, India \\
              \email{shashank.gairola@iiap.res.in}
         \and
            Pondicherry University, R.V. Nagar, Kalapet, 605014, Puducherry, India 
        \and
            Department of Physics, Indian Institute of Science Education and Research, Tirupati, Yerpedu, Tirupati - 517619, Andhra Pradesh, India
         \and
            Department of Physics and Astronomy, Rutgers University, 136 Frelinghuysen Road, Piscataway, NJ 08854, USA
        \and
            Center for Computational Astrophysics, Flatiron Institute, 162 5th Avenue, New York, NY 10010, USA
        \and
           Academia Sinica Institute of Astronomy and Astrophysics (ASIAA), No. 1, Section 4, Roosevelt Road, Taipei 10617, Taiwan
        \and
           Inter-University Centre for Astronomy and Astrophysics, Ganeshkhind, Post Bag 4, Pune 411007, India
             }

    \date{Received; accepted}

\begin{abstract}
{  
Molecular clouds fragment under the action of supersonic turbulence and gravity, which results in a scale-free hierarchical distribution of star formation within galaxies. Recent studies suggest that the hierarchical distribution of star formation in nearby galaxies shows a dependence on host galaxy properties. In this context, we study the hierarchical distribution of star formation from a few tens of parsecs up to several kiloparsecs in four nearby spiral galaxies: NGC 1566, NGC 5194, NGC 5457, and NGC 7793, by leveraging large-field-of-view and high-resolution far-ultraviolet (FUV) and near-ultraviolet (NUV) observations from the UltraViolet Imaging Telescope (UVIT). Using the two-point correlation function, we infer that the young star-forming clumps (SFCs) in the galaxies are arranged in a fractal-like hierarchical distribution, but only up to a maximum scale. This largest scale of hierarchy ($l_{\rm{corr}}$) is ubiquitous in all four galaxies and ranges from 0.5 kpc to 3.1 kpc. The flocculent spiral NGC 7793 has roughly five times smaller $l_{\rm{corr}}$ than the other three grand design spirals, possibly due to its lower mass, lower pressure environment, and a lack of strong spiral arms. $l_{\rm{corr}}$ being much smaller than the galaxy size suggests that the star formation hierarchy does not extend to the full galaxy size and it is likely an effect set by multiple physical mechanisms in the galaxy. The hierarchical distribution of SFCs dissipates almost completely within 10$-$50 Myr in our galaxy sample, signifying the migration of SFCs away from their birthplaces with increasing age. The fractal dimension of the hierarchy for our galaxies is found to be between 1.05 and 1.50. We also find that depending upon the star formation environment, significant variations can exist in the local and global hierarchy parameters of a galaxy. Overall, our results suggest that the global hierarchical properties of star formation in galaxies are not universal. This study also demonstrates the capabilities of UVIT in characterising the star formation hierarchy in nearby galaxies. In the future, a bigger sample can be employed to better understand the role of large-scale galaxy properties such as morphology and environment as well as physical processes like feedback, turbulence, shear, and interstellar medium conditions in determining the non-universal hierarchical properties of star formation in galaxies.  
}


   \maketitle

\end{abstract}

\section{Introduction}
\label{s_intro}
Star formation in galaxies is predominantly distributed in hierarchical aggregates (\citealt{2003ARA&A..41...57L}). Density peaks in the molecular clouds are the sites of the star formation process and the molecular clouds themselves are known to be hierarchically structured (\citealt{1996ApJ...471..816E}) owing to the scale-free, supersonic turbulence within them (\citealt{1981MNRAS.194..809L}).  During the early stages of star formation, the turbulent flows within the molecular clouds coupled with the gravitational Jeans instability cause the molecular clouds to fragment in a process known as gravo-turbulent fragmentation (\citealt{Elmegreen_1993}; \citealt{Padoan_2014}; \citealt{Federrath_2018}). The supersonic turbulence in the molecular clouds creates over-dense seed regions over a wide range of spatial scales (\citealt{2009ApJ...692..364F}; \citealt{2013MNRAS.436.1245F}) and through collapse followed by accretion, the seed regions grow into stars, star clusters, and star-forming complexes, which results in a scale-free hierarchical distribution of star formation within galaxies. Star formation inherits the hierarchical properties of the natal molecular clouds. This inheritance has been suggested because in the short (10-30 Myr) lifetimes of the molecular clouds (\citealt{2009ApJS..184....1K}; \citealt{Chevance_2020}; \citealt{2015ApJ...806...72M}), they share co-spatiality with the young star-forming regions of a galaxy (\citealt{Zhang_2001}; \citealt{Grasha_2018, Grasha_2019}; \citealt{2022MNRAS.516.4612T}).\

The hierarchical nature of star formation has been verified by the power-law form of the mass and size distribution function of molecular clouds  (\citealt{1996ApJ...471..816E}; \citealt{2010ApJ...720..541S}), size distribution function of star-forming complexes (\citealt{Elmegreen_2006, Elmegreen_2014}; \citealt{Gouliermis_2017}), and contour-based density analysis of young stellar groupings (\citealt{Sun_2017,2018ApJ...858...31S}; \citealt{2018AJ....156..109M}; \citealt{Rodriguez_2020}). The scale-free clustering between individual components of the star formation hierarchy such as HII regions (\citealt{Sanchez_2008}), molecular clouds (\citealt{2010ApJ...720..541S}; \citealt{Grasha_2018,Grasha_2019}), and star clusters  (\citealt{Zhang_2001}; \citealt{Gouliermis_2017}; \citealt{Grasha_2017_Spatial}; \citealt{2021MNRAS.507.5542M} (M21 from here on)) in a large number of extra-galactic systems strongly indicates that the process of star formation is hierarchical everywhere in the Universe.\

The hierarchical distribution of star formation in a galaxy shares a close resemblance with geometric fractals (\citealt{Mandelbrot_1982}). Like fractals, the clustering of the constituents of the hierarchy falls off in a power law as a function of spatial scale (\citealt{Sanchez_2008}; \citealt{Grasha_2017_Spatial}). The clustering of star clusters in particular has been shown to have a characteristic maximum scale in some of the galaxies called the correlation length ($l_{\rm{corr}}$) below which the star clusters show power-law spatial correlation with each other but on scales greater than $l_{\rm{corr}}$, the distribution of star clusters shows signatures of a nearly random distribution (\citealt{Grasha_2017_Spatial}; M21). The initial fractal-like distribution of star clusters progressively randomises with increasing age under the influence of several phenomena such as cluster drift, galactic shear, feedback, merging, and random motions. This randomisation takes the star clusters away from their parent giant molecular clouds (GMCs) over timescales of $\sim$40-100 Myr and leads to a complete loss of hierarchical signatures of star formation beyond a certain age(\citealt{Zhang_2001}; \citealt{Grasha_2017_Spatial}). This migration of star clusters away from their birthplace results in an age difference -- separation ($\Delta t - R$) relationship. For a turbulence-dominated star formation hierarchy, $\Delta t \propto R^{ 0.5}$ is expected and it was verified in both simulations \citep{2001AJ....121.1024N} and observations (\citealt{2009ApJ...700..436D,2017ApJ...842...25G}). The existence of a correlation length along with the evidence for turbulence-driven star formation process suggests that on spatial scales smaller than the correlation length, turbulence in the interstellar medium (ISM) plays a key role in setting the hierarchical distribution of star formation. Gravitational instabilities including bars and spiral arms, galactic rotation, and stellar feedback are some of the major sources that generate turbulence over a wide range of spatial scales in the ISM (\citealt{2004ARA&A..42..211E}; \citealt{2004RvMP...76..125M}; \citealt{2018PhT....71f..38F}; \citealt{2020MNRAS.493.4643M}; \citealt{2021PASP..133j2001B}; \citealt{2021NatAs...5..365F}). In particular, N-body + hydrodynamic simulations by \citet{2017MNRAS.466.1093G} show that turbulence due to stellar feedback can produce correlated density and velocity structures up to kiloparsec scales, which matches the measured order of the correlation length for some of the galaxies (\citealt{Grasha_2017_Spatial}; M21). But beyond the correlation length, turbulence can no longer maintain a scale-free hierarchical distribution of star formation.\

The hierarchical distribution of different star formation tracers in a galaxy can be quantified using a few simple parameters such as $l_{\rm{corr}}$ (the largest spatial scale up to which the distribution shows hierarchical structuring) and the fractal dimension ($D_2$ or $D_3$: the space-filling factor of the distribution, where $D_2$ is the projected fractal dimension of a three-dimensional (3D) distribution). The hierarchical structuring of molecular clouds has often been quantified using $D_3$ and they are found to have a fairly universal $D_3$ value of 2.3 $\pm$ 0.3 (\citealt{1996ApJ...471..816E}; \citealt{2010ApJ...720..541S}; \citealt{Shadmehri_2011}). In contrast, the hierarchical properties of star formation show significant deviations from universality with statistically significant variations in $D_2$ and a wide range of $l_{\rm{corr}}$ emerging in different galaxies (\citealt{Sanchez_2008}; \citealt{Grasha_2017_Spatial}; M21). These studies suggest that the hierarchy of star formation is strongly affected by physical processes such as feedback, turbulence, shear, and ambient pressure as well as galaxy morphology and environment. Comparing the hierarchy parameters for a large number of galaxies can be used to further investigate the observed non-universality in the hierarchical properties of star formation and help us better understand the physical mechanisms that impose a dependence of star formation hierarchies on galaxy morphology and environment.\

Recently, M21 used the angular two-point correlation function (TPCF) to study the hierarchical distribution of star clusters in 12 galaxies of varied morphologies. The star clusters were identified in the Hubble Space Telescope (HST) observations of galaxies under the Legacy ExtraGalactic UV Survey (LEGUS) (\citealt{2015AJ....149...51C}). Young star clusters (< 10 Myr), presumably still associated with their parent molecular clouds, were found to be hierarchically distributed up to $l_{\rm{corr}}$, which ranges from 100 pc to $>$2.5 kpc. The hierarchical nature of the star cluster distribution was found to dissipate with age, signifying the migration of star clusters away from their birthplace. In their galaxy sample, M21 found evidence for non-universality in the hierarchical properties of star formation. Moreover, the parameters describing the star formation hierarchy in the galaxies showed moderate correlation with the large-scale galaxy properties such as star formation rate density, stellar mass, and Toomre length.\

\begin{figure*}
      \centering
	   \begin{subfigure}{0.45\linewidth}
		\includegraphics[width=\linewidth]{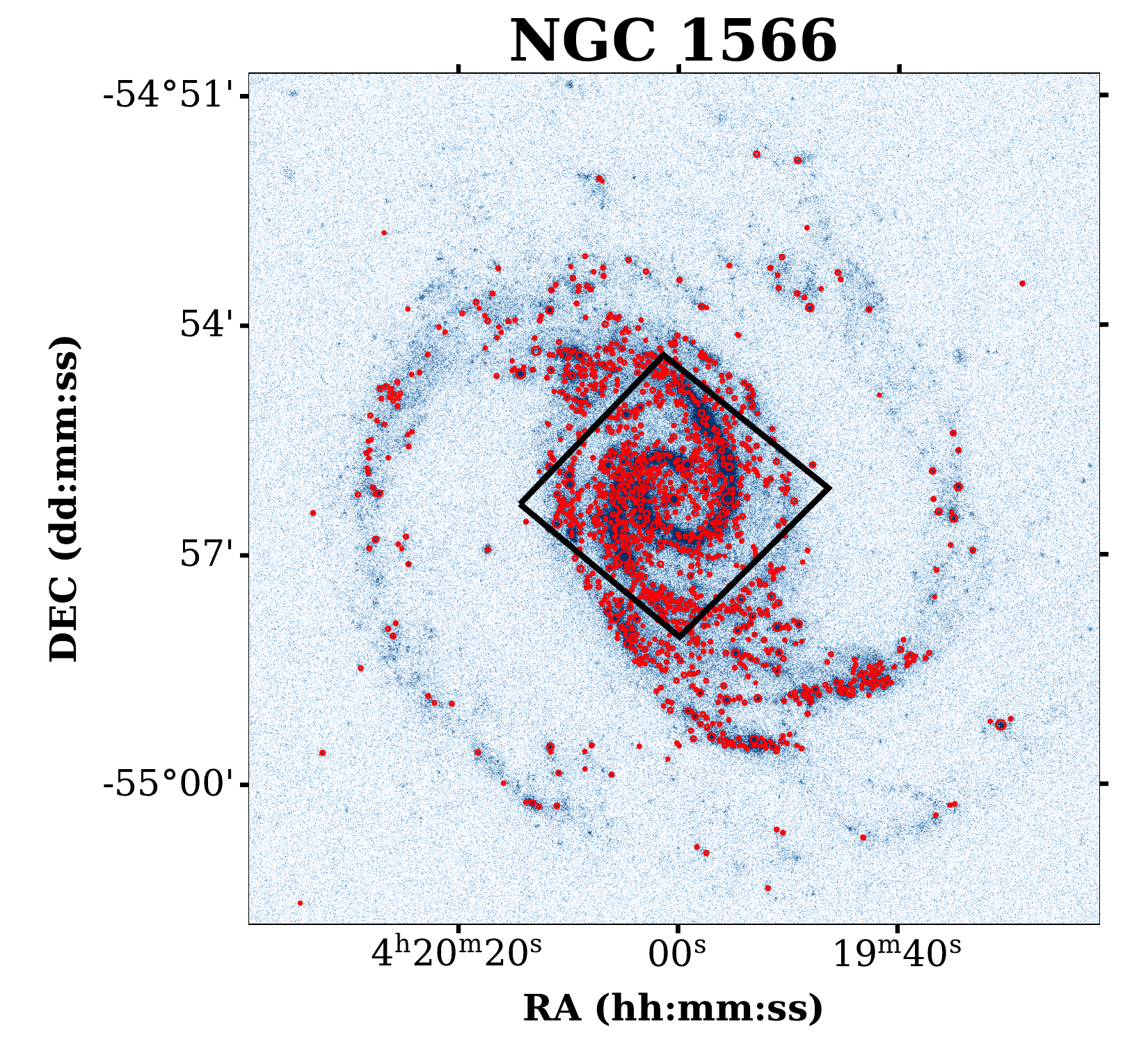}
		\label{fig:subfig1}
	   \end{subfigure}
	   \begin{subfigure}{0.45\linewidth}
		\includegraphics[width=\linewidth]{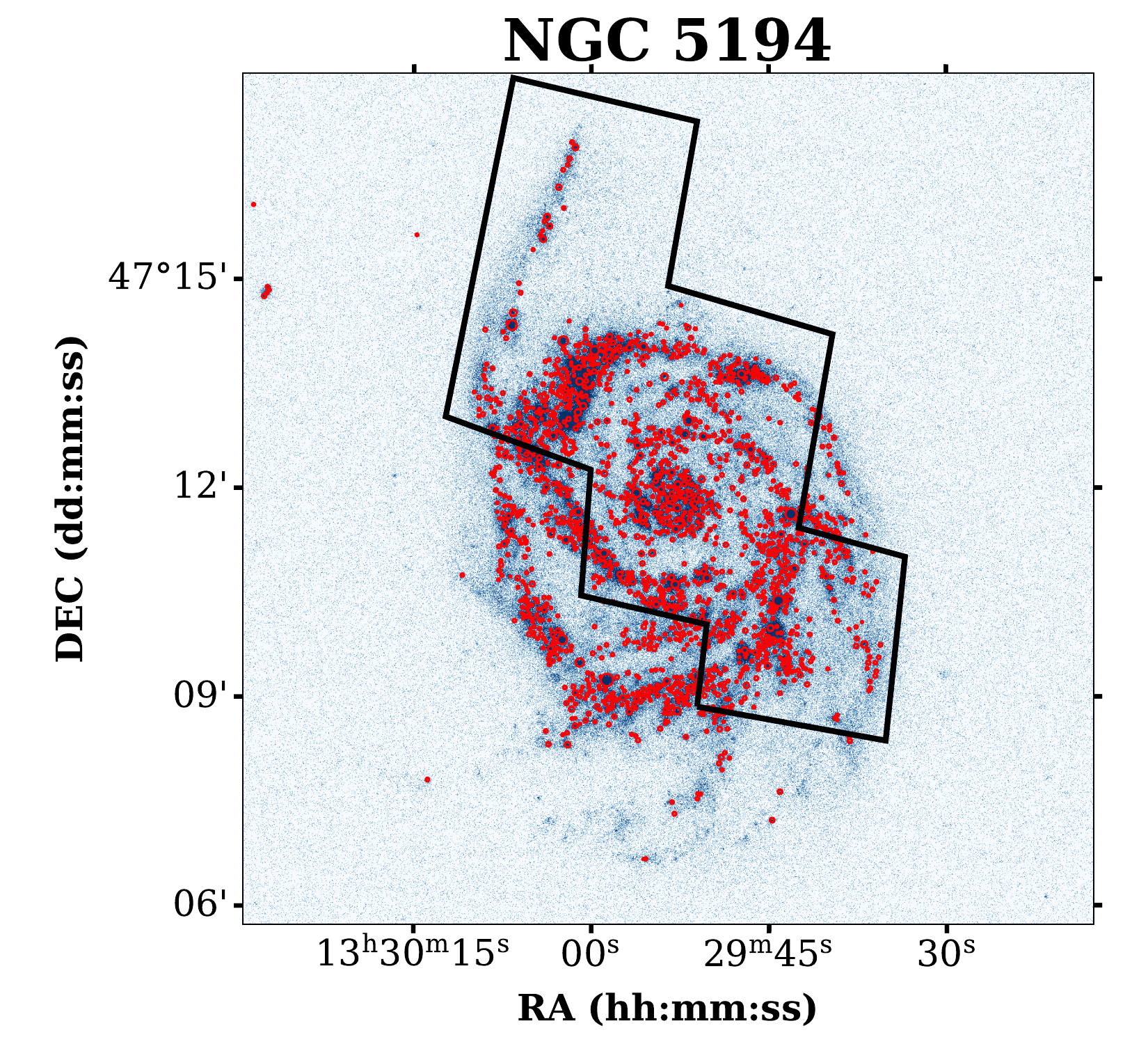}
		\label{fig:subfig2}
	    \end{subfigure}
	\vfill
	     \begin{subfigure}{0.45\linewidth}
		 \includegraphics[width=\linewidth]{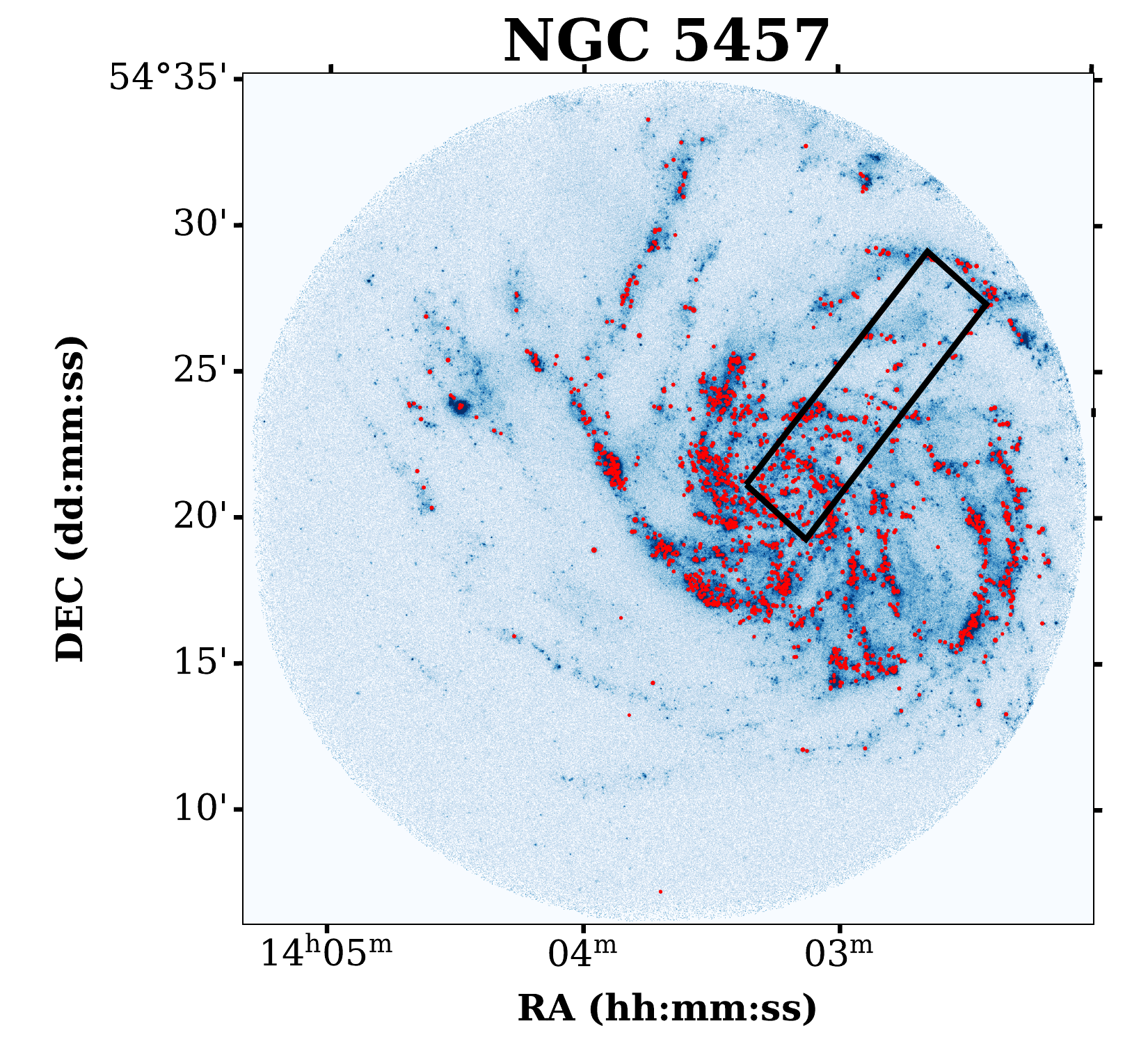}
		 \label{fig:subfig3}
	      \end{subfigure}
	       \begin{subfigure}{0.45\linewidth}
		  \includegraphics[width=\linewidth]{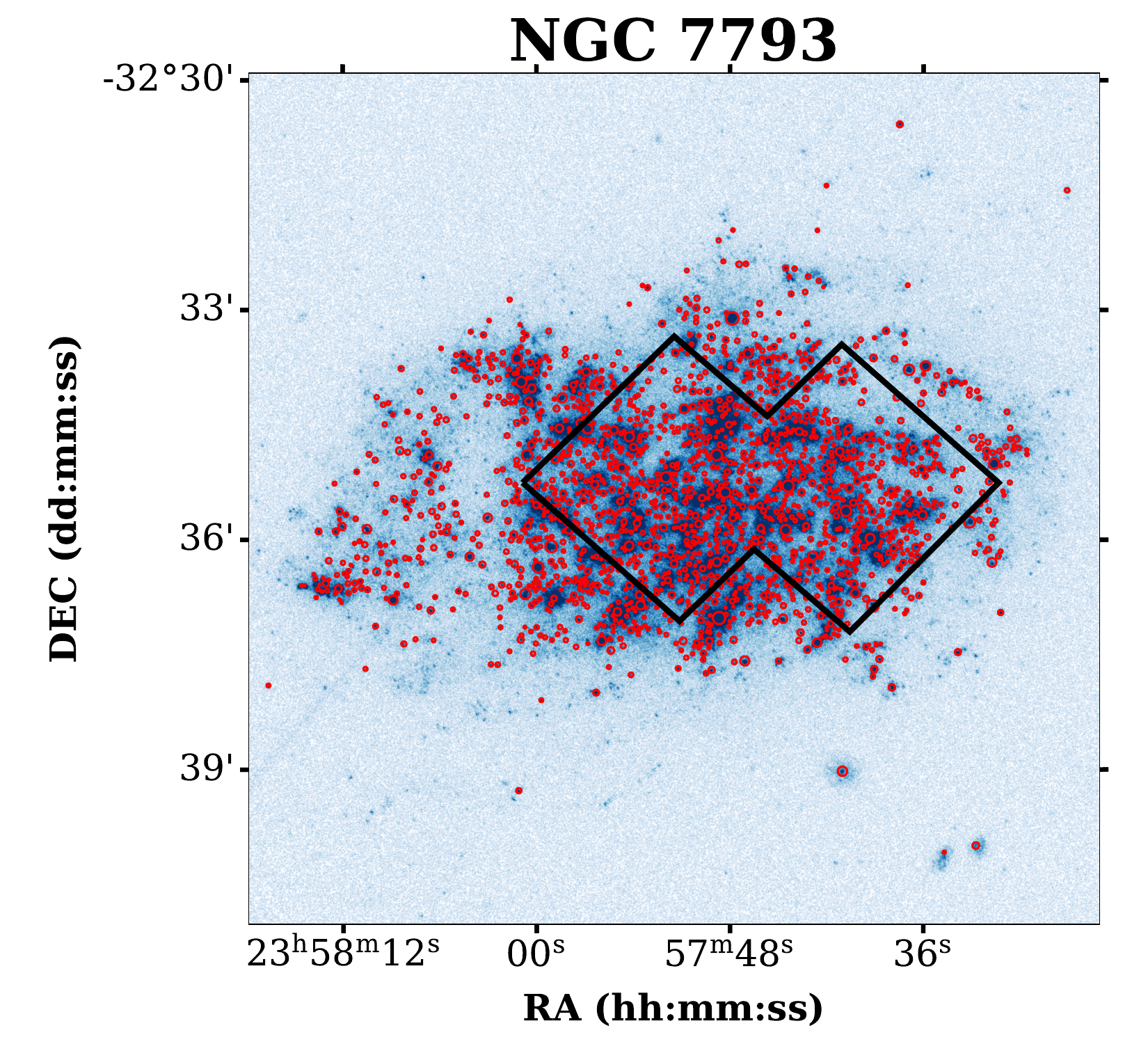}
		  \label{fig:subfig4}
	       \end{subfigure}
\caption{UVIT FUV images of the four galaxies in our sample. The red circles represent the locations and sizes of the SFCs in the galaxies that were detected using astrodendro and the black polygons represent part of the galaxy area that was previously observed using the HST and used in M21. Due to its large FoV, UVIT is able to cover the entire galaxy and it provides an advantage over an instrument like the HST.}
\label{galaxy_images}
\end{figure*}

For some of their sample galaxies like NGC 1566 and NGC 5194, M21 found that the power-law spatial correlation of star clusters exists till the largest measurable spatial separation between star clusters. This made it appear as if the turbulence-driven star formation hierarchy in these galaxies could extend up to length scales of the order of the entire galaxy size. This raises the interesting question of whether $l_{\rm{corr}}$ even exists in these galaxies or whether the observed spatial correlation is just capturing the signal from the large scale of the galaxy. However, these galaxies were only partially covered by the HST (See Figure \ref{galaxy_images}). So, if the actual galaxy $l_{\rm{corr}}$ is larger than or of the order of the galaxy size covered by the HST, then the spatial correlation is expected to exist till the largest measurable spatial separation between star clusters, and this length scale is just the lower limit of $l_{\rm{corr}}$. Additionally, in the case of NGC 5457, the HST coverage of the galaxy is approximately 20\%. Therefore, it is plausible that the hierarchy parameters valid for the HST-covered region of NGC 5457 can only describe the local star formation hierarchy and that the global hierarchy parameters for the galaxy might be entirely different. Full galaxy coverage would be needed to overcome these challenges and to better estimate the hierarchy parameters of the aforementioned galaxies.\

It is against this backdrop that we study hierarchical star formation and aim to reliably estimate the global hierarchy parameters for four nearby spiral galaxies (which were part of the M21 galaxy sample): NGC 1566, NGC 5194, NGC 5457, and NGC 7793. For this purpose, we have used the far-ultraviolet (FUV) and near-ultraviolet (NUV) observations of these galaxies obtained by the UltraViolet Imaging Telescope (UVIT) on board AstroSat (\citealt{2006AdSpR..38.2989A}; \citealt{2014SPIE.9144E..1SS}; \citealt{2012SPIE.8443E..1NK}). While leveraging the full galaxy coverage provided by the 28$^\prime$ field of view (FoV) of UVIT, we aim to overcome the limitations of M21 and, in the process, test the capabilities of UVIT in studying hierarchical star formation. We have based our study on the UVIT-FUV observations for their high sensitivity in tracing the young star-forming regions in a galaxy. Compared to NUV (the shortest available band in M21), FUV is less affected by the crowding of flux caused by the relatively evolved, older regions. Moreover, due to its better angular resolution (1.5$^{\prime\prime}$), UVIT offers an advantage over previous UV telescopes such as the Galaxy Evolution Explorer (GALEX). Our study will contribute extensively towards understanding the role of turbulence, galaxy environment, and morphology in the star formation process and provide valuable insights about the non-universality of hierarchical star formation.\

The remaining paper is structured as follows. The galaxies chosen for this study and the data used are presented in Sections \ref{s_sample} and \ref{s_data}, respectively. In Section \ref{s_analysis}, we explain our methodology, and in Section \ref{ss_models}, we present the mathematical models that are used to describe the observed TPCF. In Section \ref{s_results}, we describe our results, and Section \ref{s_discussion} provides a discussion of our findings. Finally, in Section \ref{s_summary}, we summarise our results and provide a brief plan for a follow-up study in future.\

\section{Galaxy sample}
Our galaxy sample consists of NGC 1566, NGC 5194, NGC 5457, and NGC 7793, chosen from the galaxy sample of M21. For these galaxies, we provide some general galaxy properties used in this paper in Table \ref{table1}.\

\begin{table*} 
\caption{Some important properties of the selected galaxies in this study.}
\centering
\begin{tabular}{cccccccc}
\hline
Galaxy & Morphology & Distance (Mpc) & Position angle (deg) & Inclination angle (deg) & Stellar Mass ($M_{\odot}$)\\
\hline
NGC 1566 & SABbc & 17.7 & 214.7 & 29.6 & 2.7 $\times$ 10$^{10}$\\
NGC 5194 & SAbc & 8.6 & 173.0 & 22.0 & 2.4 $\times$ 10$^{10}$\\
NGC 5457 & SABcd & 6.7 & 39.0 & 18.0 & 1.9 $\times$ 10$^{10}$\\
NGC 7793 & SAd & 3.6 & 98.0 & 55.0 & 3.2 $\times$ 10$^{9}$\\
\hline
\end{tabular}
\tablefoot{Name, morphology (\citealt{deVaucouleurs_1991}), distance, position angle, inclination angle and stellar mass of the galaxies are taken from the Table 1 of \citet{2021MNRAS.507.5542M}.}
\label{table1}
\end{table*}
\label{s_sample}

\begin{table*}
\caption{Description of the UVIT observations of the sample galaxies and the number of SFCs identified in each galaxy.}
\centering
\begin{tabular}
{p{0.80in}p{1.2in}p{0.40in}p{0.40in}p{1.20in}p{0.40in}p{0.40in}p{0.50in}}
\hline
Galaxy \newline \newline (1) & FUV filter : \newline Exposure time \newline (2) & $\lambda_{FUV}$ ({\AA}) \newline (3) & $\Delta \lambda_{FUV}$ ({\AA}) \newline (4) &  NUV filter : \newline Exposure time \newline (5) & $\lambda_{NUV}$ ({\AA}) \newline (6) & $\Delta \lambda_{NUV}$ ({\AA}) \newline (7)  & $N_{SFC}$ \newline \newline (8) \\\hline
NGC 1566 & F148W : 2940 sec & 1481 & 500 & N263M : 2964 sec & 2632 & 275 & 1385\\
NGC 5194 & F148W : 1882 sec & 1481 & 500 & N263M : 1024 sec & 2632 & 275 & 1912\\
NGC 5457 & F148W : 3163 sec & 1481 & 500 & N263M : 2987 sec & 2632 & 275 & 1548\\
NGC 7793 & F148W : 7565 sec & 1481 & 500 & N242W : 8108 sec & 2418 & 785 & 1931\\\hline
\end{tabular}
\tablefoot{ (1): Galaxy name. (2): UVIT FUV filter used and the FUV exposure time. (3): Central wavelength of the FUV filter. (4): Bandwidth of the FUV filter. (5): UVIT NUV filter used and the NUV exposure time. (6): Central wavelength of the NUV filter. (7): Bandwidth of the NUV filter. (8): Total number of SFCs found in each galaxy after applying FUV and NUV magnitude error cuts of 0.20 in NGC 1566 and NGC 5194 and 0.10 in NGC 5457 and NGC 7793 (see Section \ref{ss_characterization}.2 for more details).}
\label{table2}
\end{table*}

Due to their large angular sizes, the star-forming disc of these galaxies was only partially covered (20-70\%) by the HST, even with multiple pointings. The difference in galaxy area covered by the HST and the UVIT observations can be clearly seen from Fig. \ref{galaxy_images}. NGC 1566 and NGC 5194 are chosen because for these galaxies only a lower limit for $l_{\rm{corr}}$ could be found by M21 (Section \ref{s_intro}). They derived a lower limit of 1.7 kpc and 2.7 kpc for $l_{\rm{corr}}$ and computed the fractal dimension ($D_2$) to be 1.5 and 1.6 in NGC 1566 and NGC 5194, respectively. NGC 5457 had the least amount of galaxy coverage (< 20\%) in M21 so, the global hierarchy parameters for NGC 5457 have not been properly constrained, to date. Moreover, it is an ideal galaxy for investigating how the hierarchy parameters change with the amount of galaxy coverage. Based on the data from the partial coverage of NGC 5457, M21 found its $l_{\rm{corr}}$ and $D_{2}$ to be $\sim$450 pc and 1.4, respectively. Finally, to compare the hierarchy parameters derived using the HST and the UVIT data, we also need a reference galaxy that was mostly covered by the HST. We choose NGC 7793 for this purpose for which M21 derived $l_{\rm{corr}}$ and $D_{2}$ to be $\sim$100 pc and 0.5, respectively.\

Since the hierarchical distribution of star formation in a galaxy dissipates with age, distinguishing between young and old star-forming regions is essential. We used the available archival FUV and NUV observations of the sample galaxies taken from the UVIT for this purpose (Section \ref{ss_age}.3).\

\section{Data}
\label{s_data}
The UV emission arising from the galaxies can be used as a direct tracer of recent ($\sim$ 100 Myr) star formation (\citealt{2012ARA&A..50..531K}). In order to study the star-forming regions in the galaxies, we used archival FUV and NUV UVIT observations. UVIT is one of the payloads on board AstroSat, which is India's first space-based multi-wavelength observatory (\citealt{2012SPIE.8443E..1NK}). It is a high resolution ($\sim$ 1.5$^{\prime\prime}$) imaging instrument with a large, circular FoV of $\sim$ 28$^\prime$ diameter. It offers simultaneous FUV (1300-1800 \AA) and NUV (1800-3000 \AA) imaging for science objectives and the visible (VIS : 3200-5500 \AA) channel of the UVIT is used to track the spacecraft drift during observations.\

We used the UVIT data reduction software CCDLAB (\citealt{2021JApA...42...30P}) to create the science-ready images of the galaxies. Different orbit-wise images were first cleaned for charge-coupled device (CCD) artefacts such as cosmic ray hits and using the VIS channel observations, these images were corrected for the low-frequency drift experienced by the spacecraft during an orbit of observation. These images were then corrected for flat-field and distortions using the calibrations provided in \citet{2020AJ....159..158T}. Different orbit-wise images were subsequently aligned on top of each other using a registration procedure and merged to produce the galaxy image corresponding to a given filter. Then, the point spread function (PSF) was optimised for the filter-wise images, which typically results in a PSF-full width half maximum (FWHM) of < 1.5$^{\prime\prime}$. The PSF optimisation step is crucial because it removes any residual drift left over from the drift correction and provides a consistent PSF across the full 28$^\prime$ FoV of the UVIT. Lastly, astrometry was performed on the science images by matching bright sources in the UVIT field with the Gaia Data Release 3 catalogue to establish the world coordinate system in the image. A summary of the UVIT data used in this study is given in Table \ref{table2}.\

\section{Analysis} 
\label{s_analysis}
\textbf{\subsection{Identification of star-forming clumps}}
\label{ss_identification}
The FUV emission coming from star-forming galaxies is known to be clumpy. The UVIT resolution ($\sim$ 70 pc at 10 Mpc distance) only allows us to detect large clumps of star formation. Physically, each clump is probably an OB association or a collection of star clusters that are recently formed ($\lesssim$ 100 Myr). We used the Astrodendro\footnote{\url{https://dendrograms.readthedocs.io/en/stable/}} package (\citealt{2008ApJ...679.1338R}) in Python to identify star-forming clumps (SFCs) in the FUV galaxy images. Astrodendro identifies local flux maximas in an intensity map (such as a fits image of a galaxy) and generates a dendrogram tree of the identified structures. The structures in a dendrogram share parent-child relationships and are classified in the form of leaves, branches and trunks. Leaves are the smallest and densest flux structures found in a dendrogram and they have no further sub-structures or children. This makes the leaves ideal candidates to be classified as unresolved SFCs. A collection of leaves that are inter-connected with each other constitutes a branch in the dendrogram. The branches are the parent structures of the leaves and they are typically much bigger and less dense than leaves. Branches themselves are hosted by the dendrogram trunk, which is the largest and least dense structure possible in a dendrogram. The trunk has no parent structure of its own and it is considered a parent to the branches and isolated leaves. This parent-child relationship between different flux-structures makes the Astrodendro package extremely well suited to explore the hierarchical structuring of UV emission in a galaxy where brighter star-forming regions are embedded within progressively fainter regions across a wide range of spatial scales.\

Astrodendro uses three parameters -- min\_value, min\_delta, and min\_npix -- to detect SFCs. In our analysis, SFCs were identified in the FUV image because FUV emission can best probe the young stellar population. The min\_value parameter was used to set the threshold flux above which all the SFCs should be identified. We used the background + three times the standard deviation of the background flux (1 bg + 3 $\sigma$) as the min\_value. The 1 bg in min\_value takes into account the UV sky background.  Effectively, we are detecting SFCs on a background subtracted FUV image. The bg and $\sigma$ values for each galaxy were calculated by placing eight circular apertures of 1 arcminute size each on the observed UVIT field containing the galaxy. These apertures were placed far away from the galaxy and other bright sources. The flux from these eight apertures was then averaged to get the final bg and $\sigma$ values, which we have used in the dendrogram input parameters. The min\_delta parameter was used to avoid spurious clump detection caused by random noise adding on top of the local flux average and posing as a distinct SFC. We chose 1 $\sigma$ as the min\_delta parameter. The min\_npix parameter specifies the minimum number of connected pixels (each having flux value > min\_value) an identified structure should have. Because the SFCs are expected to be extended sources, a value of 11 pixels was used as min\_npix, which roughly corresponds to the average area contained within the UVIT PSF. After the dendrogram was constructed from the galaxy's FUV image using the parameters described above, all the resulting leaves in the dendrogram, which are the local peaks of FUV emission, were taken as individual SFCs in our analysis.\

Within a galaxy, there also exists a local UV background that arises from sources such as: 1) low-density, young star-forming regions that remain undetected as SFCs; and 2) evolved, hot stellar populations such as the horizontal branch stars, post-asymptotic-giant-branch stars, and white dwarfs (\citealt{2021JApA...42...85D}). Though our Astrodendro detection threshold (min\_value = 1 bg + 3 $\sigma$) is based not on the local UV background but rather on the bg and $\sigma$ values measured from the sky regions, we found that the effect of local UV background on the SFC sizes is negligible (see Appendix \ref{appdx5} for more details). Lastly, as our investigation of hierarchical star formation is going to be based on the statistical method of spatial correlation among SFCs that is independent of SFC sizes, our final results are unlikely to be influenced by the local UV background.\

We found that in all four galaxies, if we equate the area of the observed SFCs to a circle in the sky plane, then the SFC sizes range from 1.5$^{\prime\prime}$ to 8.5$^{\prime\prime}$ diameter. This roughly corresponds to SFC diameters of 25 - 150 pc, 50 - 280 pc, 60 - 350 pc, and 130 - 730 pc at the galaxy distance for NGC 7793, NGC 5457, NGC 5194, and NGC 1566, respectively.\

\textbf{\subsection{Photometric characterisation of star-forming clumps}}
\label{ss_characterization}
Since we aimed to derive SFC ages using (FUV$-$NUV) colour (Section \ref{ss_age}.3), we needed the FUV and NUV magnitudes for all the SFCs. To achieve this, we adopted the same procedure as \citet{Mondal_2021}. Because for UVIT, the NUV angular resolution ($<$ 1.2$^{\prime\prime}$) is typically much better than the FUV angular resolution ($<$ 1.5$^{\prime\prime}$), we degraded the NUV images to match the FUV image resolution. The resulting, common resolution of FUV and NUV images was found to be better than 1.5$^{\prime\prime}$, which roughly corresponds to 25 $-$ 130 pc at the distance of the galaxies in our sample. Then, for all the SFCs, we created circular apertures with each aperture having the same position and total area as the corresponding SFC identified in the FUV image by Astrodendro. These circular apertures were placed at the positions of the astrodendro-identified SFCs and the FUV and NUV fluxes for all the SFCs were extracted using the photutils package in Python. The sky background was also subtracted from the extracted FUV and NUV fluxes. Finally, the extracted fluxes were converted to apparent AB magnitudes using the calibration relations of \citet{2020AJ....159..158T}.\

We corrected the estimated FUV and NUV magnitudes of all the SFCs for interstellar extinction. For NGC 7793, due to its Large Magellanic Cloud (LMC) -like metallicity (\citealt{2012AJ....143...19V}), the LMC extinction law was used by \citet{Mondal_2021}. They used an E(B$-$V) value of 0.179 mag, taken from \citet{2010MNRAS.405.2737B}. Since we used the same UVIT data as \citet{Mondal_2021}, we directly adopted the FUV and NUV extinction correction ($A_{FUV}$ = 1.75 mag and $A_{NUV}$ = 1.45 mag) from their paper, which takes into account both the Milky Way and the host galaxy extinction. For the remaining three galaxies, the $A_{\lambda}$ values for the UVIT filters were calculated using Cardelli extinction law (\citealt{1989ApJ...345..245C}). The Milky Way extinction values were taken from the dust maps of \citet{2011ApJ...737..103S}. The internal extinction values for these galaxies were taken from recent papers all of which used spectral energy distribution fitting of LEGUS star clusters to derive the extinction values (NGC 1566 : $A_V$ = 0.55 mag (\citealt{Gouliermis_2017}) , NGC 5194 :  $A_V$ = 1.0 mag (\citealt{Grasha_2018}), NGC 5457 : $A_V$ = 0.74 mag (\citealt{2022ApJ...935..166L})). For NGC 1566 and NGC 5457, Milky Way attenuation curve whereas for NGC 5194, the starburst attenuation curve was used in the aforementioned papers. The computed ($A_{FUV}$, $A_{NUV}$) values for the UVIT filters used (Table \ref{table2}) in NGC 1566, NGC 5194 and NGC 5457 are (1.48, 1.16), (2.93, 2.30) and (2.10, 1.65) mag, respectively.\

For further analysis, we introduced magnitude error cuts of 0.10 and 0.20 on both FUV and NUV magnitudes, so that only significant detections of SFCs are considered in our analysis. This helps in minimising the errors associated with the derived SFC ages in section \ref{ss_age}.3. For NGC 5457 and NGC 7793, because of high-quality data, we used SFCs with magnitude errors of less than 0.10. However, for NGC 1566 and NGC 5194, we detected a much smaller number of SFCs with magnitude errors of less than 0.10. This is due to a combination of lower exposure times and comparatively large distances for NGC 1566 and NGC 5194. So, for NGC 1566 and NGC 5194, we used SFCs with magnitude errors of less than 0.20 in our analysis. For each galaxy, the final number of SFCs used in our analysis after the magnitude cuts were applied is given in Table \ref{table2}.\

\textbf{\subsection{Age estimation of the star-forming clumps using Starburst99}}
\label{ss_age}
Since the hierarchical distribution of star-forming regions is known to undergo significant age evolution, it is essential to distinguish young regions from the old ones. Star-forming regions rich in OB stars, emit strongly in FUV and NUV. As the star-forming regions age, the most massive OB stars within them begin to disappear. This results in a gradual decline of the UV spectral slope ($\beta_{UV}$ : the ratio of the FUV flux compared to the NUV flux arising from the star-forming region) and consequently the (FUV$-$NUV) colour of any star-forming region moves to redder values. But, the redness of a star-forming region can also be due to dust extinction therefore, it is crucial to break this extinction-age degeneracy of the (FUV$-$NUV) colour.\

With extinction values derived from sources apart from $\beta_{UV}$, the extinction-corrected (FUV$-$NUV) colour can be used for age estimation. To achieve this, we generated a synthetic colour-age relationship using simple, single-age stellar populations in the Starburst99 (SB99) simulation (\citealt{1999ApJS..123....3L}). SB99 simulates a star cluster for input parameters such as total stellar mass, metallicity, initial mass function (IMF) and evolutionary tracks. For our analysis, we have compared the properties of the star clusters derived using SB99 with the observed SFC properties. SB99 takes in the assumption that any observed SFC  fully samples the IMF, which is approximately true for SFCs of mass greater than $10^3 M_{\odot}$. All the star formation within an SFC is assumed to have happened in an instantaneous burst. The details of the input parameters used in SB99 are provided in Table \ref{table3}. Based on the oxygen abundance in NGC 7793, \citet{2012AJ....143...19V} reported a metallicity value of Z $\sim$ 0.012 so, we used Z = 0.008 which is the closest allowed value in SB99. For the remaining three galaxies, since their stellar masses are comparable to that of the Milky Way, we used Z = 0.02, which roughly corresponds to the solar metallicity (\citealt{2009ARA&A..47..481A}).\

\begin{table}
\caption{Important input parameters used in Starburst99}
\centering
\resizebox{90mm}{!}{
\begin{tabular}{cc}
\hline
Parameter & Value\\\hline
Star formation  type & Instantaneous\\
Initial mass function (IMF) & Kroupa [1.3, 2.3]\\
Stellar mass range & 0.1, 0.5, 120 $M_{\odot}$\\
Clump mass & $10^3 M_{\odot}$-$10^6 M_{\odot}$\\
Metallicity & Z=0.008 (NGC 7793) \\
& Z=0.02(remaining 3 galaxies)\\
Evolutionary track & Geneva (high mass loss rate)\\
Age range & 1-900 Myr\\\hline
\end{tabular}}
\label{table3}
\end{table}

The output spectra from the SB99 gives the total energy output for an individual, unresolved SFC of a given mass and age. We convolved these spectra with the UVIT filter transmission curves (\citealt{2020AJ....159..158T}) and normalised the flux with the total effective area of the filter. Using the distance of each galaxy, we converted the flux received through the FUV and NUV filters to apparent AB magnitudes and generated the synthetic colour-magnitude diagram (CMD) like the one presented in Figure \ref{cmd} for NGC 7793. Similar CMDs generated for NGC 1566, NGC 5194 and NGC 5457 are presented in Appendix \ref{appdx3}. As was expected, the SFCs with higher masses are brighter and SFCs get progressively redder with age. We compared the CMD of the observed SFCs with the synthetic SB99 CMD for our sample galaxies and interpolated along the colour axis to calculate the age of each SFC identified in the sample galaxies.\

We propagated the errors in FUV$-$NUV colour to derive the errors associated with the SFC ages. For NGC 7793 and NGC 5457, where 0.10 magnitude cuts were applied to the SFCs, the median age errors are $\sim$3 Myr and $\sim$23 Myr for SFCs with ages of less than 20 Myrs and between 20$-$100 Myr, respectively. For NGC 1566 and NGC 5194, where 0.20 magnitude cuts are applied to the SFCs, the median age errors are $\sim$6 Myr and $\sim$45 Myr for SFCs with ages of less than 20 Myrs and between 20$-$100 Myr, respectively.\

\begin{figure}
    \centering
    \includegraphics[width=0.45\textwidth]{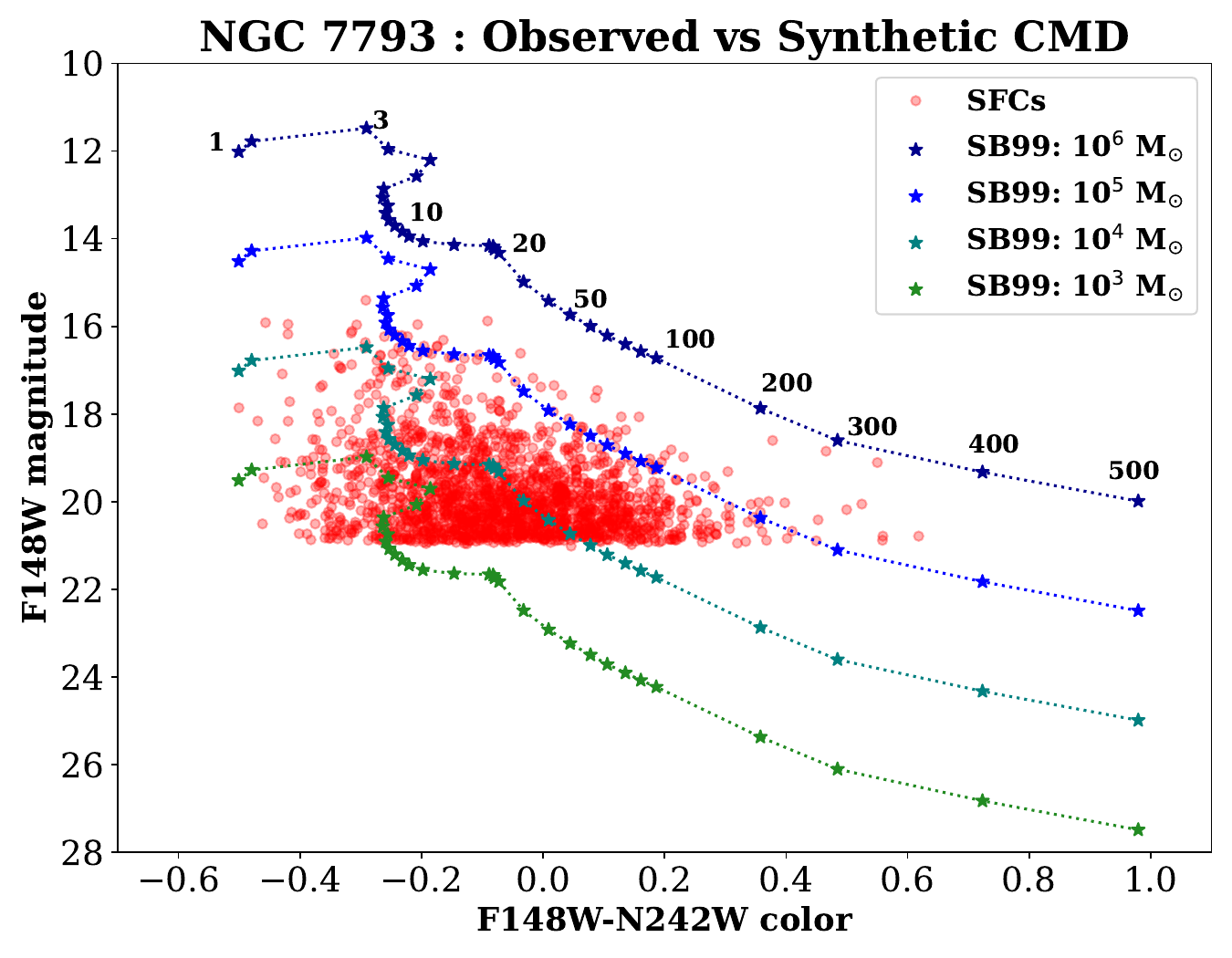}
    \caption{Synthetic colour-magnitude diagram from Starburst99 and the extinction-corrected SFCs (red dots) identified in NGC 7793. The numbers annotated above the $10^6 M_{\odot}$ SFC track represents the ages of the synthetic stellar population in Myr. The FUV$-$NUV colour can be seen getting progressively redder with age in this figure.}
    \label{cmd}
\end{figure}

\citet{Chevance_2020} found that in a sample of nine nearby star-forming galaxies, star clusters dissociate themselves from their parent molecular cloud and their birthplace on an average timescale of $\sim$5 Myr. To distinguish star clusters presumably still associated with their parent molecular clouds from the star clusters no longer associated with their parent molecular clouds, M21 used a slightly relaxed age-cut of 10 Myr. This relaxed cut was adopted due to the uncertainties in the estimated ages of the star clusters. We implemented the same age-cut of 10 Myr for our SFCs. From here on, younger than 10 Myr SFCs will be called `young' and older than 10 Myr SFCs will be called `old', unless stated otherwise explicitly. While checking how old an SFC detected in an FUV image can be, we found that in NGC 7793, more than 95\% of SFCs are less than 100 Myr old and $\sim$4\% of SFCs are 100-200 Myr old. The remaining $\sim$1\% of SFCs are older than 200 Myr. A similar age demographic for the SFCs is found in the other three galaxies too.\

\begin{figure}
    \centering
    \includegraphics[width=0.45\textwidth]{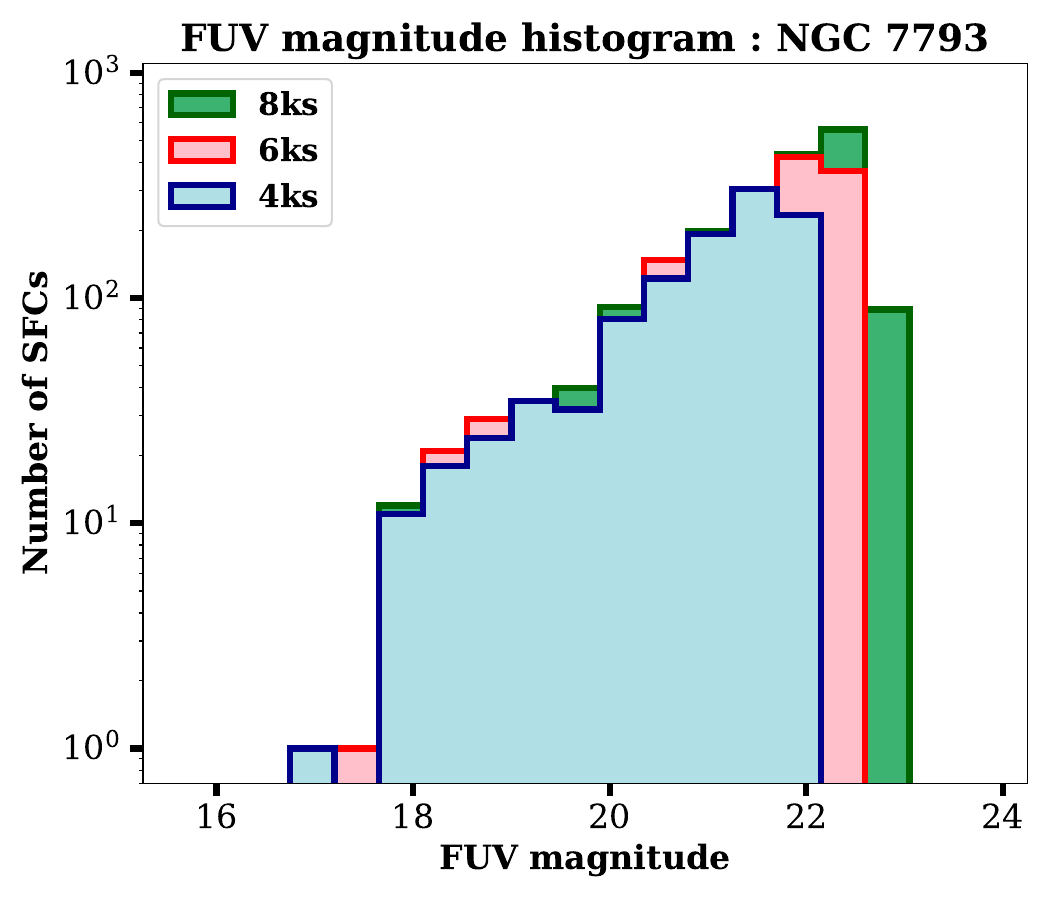}
    \caption{FUV magnitude histogram of the SFCs detected in the FUV images of NGC 7793 which have 4ks, 6ks, and 8ks exposure times. It is evident in this figure that with higher exposure times, we can detect more faint SFCs. The peak of the histogram is considered as the approximate completeness limit in our analysis. }
    \label{completeness}
\end{figure}

\textbf{\subsection{Completeness limit of the star-forming clumps}}
\label{ss_completeness}
To derive the completeness limit of our observed SFCs, we took NGC 7793 as the reference system because it has the highest exposure time in our sample, approximately 8 kiloseconds (ks). We intentionally removed some orbit-wise images from its UVIT observations to create two NGC 7793 images of 6ks and 4ks exposure time. Then, we identified the SFCs in these two images as described in Section \ref{ss_identification}.1 and also applied a magnitude error cut of 0.1 mag to limit our analysis to the significant SFC detections. Next, we plotted the FUV magnitude histograms for the SFCs detected in the 4ks, 6ks, and 8ks images of NGC 7793, which can be seen in the Figure \ref{completeness}. As was expected, the peak of the histogram progressively shifts to fainter magnitudes as the exposure time of the considered image increases. From the figure, we can see that the number of SFCs in the magnitude bin corresponding to the peak of the histogram and those in the brighter magnitude bins, detected from a shallow exposure image, remain the same when we compare with a histogram of SFCs detected from a deeper image. This suggests that the magnitude corresponding to the peak of the histogram can be considered as the completeness limit of the SFCs detected from that image (\citealt{2020ApJS..247...47L}; \citealt{2023ApJ...946...65D}; \citealt{2024AJ....168..255H}). Based on this analysis we find the FUV magnitudes corresponding to the peak of the number distribution of the detected SFCs in all the four galaxies in our sample (with respective magnitude error cuts applied as discussed in Section \ref{ss_characterization}.2) and use them as the completeness limit of the detected SFCs in each galaxy.\ 

The FUV magnitude corresponding to the completeness limit is corrected for extinction and then matched against the SB99 CMD of the galaxy. Then the age is computed for a synthetic SB99 SFC of $10^4 M_{\odot}$ and FUV magnitude equal to the extinction-corrected completeness limit. With this analysis, we found that the SFCs used for the TPCF analysis in this paper with appropriate extinction correction and magnitude error cuts applied (Section \ref{ss_characterization}.2) are approximately complete till 7, 6, 8, and 50 Myr ages in NGC 1566, NGC 5194, NGC 5457, and NGC 7793, respectively.\

\textbf{\subsection{Galaxy de-projection}}
\label{ss_deprojection}
Galaxies are usually inclined with respect to the sky plane, which means that the true spatial separation between any two SFCs in a galaxy will be larger than their observed separation. As our analysis relies on the spatial separation between SFCs, to get the true spatial separation we de-projected all the SFC positions within the galaxies under the assumption that the galaxies are relatively thin, circular discs. In equation 1, using the galaxy position angle ($\phi$; measured counter-clockwise from the celestial North), the SFC positions were de-rotated so that the galaxy major axis is aligned with the celestial North-South axis. Then, equation 2 was used to stretch the minor axis of the galaxy using the inclination angle ($i$) so that the major and minor axis of the galaxy are nearly equal, as was expected from an ideal, circular disc galaxy. As a result, X and Y represent the de-projected coordinates of SFCs.
\begin{equation}
\begin{array}{l}
x' = x_0 \cos (\phi) + y_0 \sin (\phi) \\
y' = y_0 \cos (\phi) - x_0 \sin (\phi) 
\end{array}
\end{equation}
\begin{equation}
\begin{array}{l}
X = x'/ \cos (i) \quad ;\quad \quad Y = y'
\end{array}
.\end{equation}

\textbf{\subsection{Two-point correlation function}}
\label{ss_TPCF}
In several recent studies, the TPCF has proved to be a powerful tool to study star formation hierarchies (see Section \ref{s_intro}). The TPCF quantifies the clustering of constituting elements over a wide range of spatial scales to characterise any hierarchical distribution. It signifies the amount of excess spatial correlation a distribution has over a completely random distribution occupying the same spatial extent. We used the Landy-Szalay (LS) TPCF estimator (\citealt{Landy_1993}) in our analysis, which is described in the following equations:
\begin{equation}
\begin{array}{l}
\mathrm{TPCF}(x) = 1 + w_{\mathrm{LS}}(x)
\end{array}
\end{equation}
\begin{equation}
\begin{array}{l}
w_{\mathrm{LS}}(x)=\frac{\mathrm{DD}(x)-2 \mathrm{DR}(x)+\operatorname{RR}(x)}{\operatorname{RR}(x)}
\end{array}
\end{equation}
\begin{equation}
\begin{array}{l}
\mathrm{DD}(x)=\frac{P_{\mathrm{DD}}(x)}{N_D(N_D-1)};\quad \mathrm{DR}(x)=\frac{P_{\mathrm{DR}}(x)}{N_D N_{R}};\quad \operatorname{RR}(x)=\frac{P_{\mathrm{RR}}(x)}{N_{R}\left(N_{R}-1\right)}\\
\end{array}
.\end{equation} 

For point-like distributions, the LS estimator uses pair counting to measure the clustering of data points. First, a random distribution of data points is generated, which populates the same spatial footprint as the actual data points. Then, a census of data-data ($P_{\mathrm{DD}}$), data-random ($P_{\mathrm{DR}}$), and random-random ($P_{\mathrm{RR}}$) pairs is conducted within a spatial separation bin centred at $x$. The pair counts are normalised by the number of data points ($N_D$) and random points ($N_R$) (equation 5) and the normalised pair counts (DD, DR, RR) are used in equations 3 and 4 to compute the TPCF for the given spatial separation bin. A completely random distribution has equal amount of clustering at all spatial scales, so it has TPCF = 1 and zero excess correlation. For purely hierarchical distributions, TPCF has a power-law form: TPCF($x$) = A $x$$^{-\alpha_1}$, where A is a normalisation constant and $\alpha_1$ is related to the 2D fractal dimension ($D_2$) by $D_2$ = 2 + $\alpha_1$. $D_2$ quantifies the space-filling factor of the distribution. Physically, $D_2$ provides information on the amount of clustering in the distribution of star formation within the galaxies and hints at the natal turbulent density structures from which the stars have formed (\citealt{2009ApJ...692..364F})\

\begin{figure}
    \centering
    \includegraphics[width=0.45\textwidth]{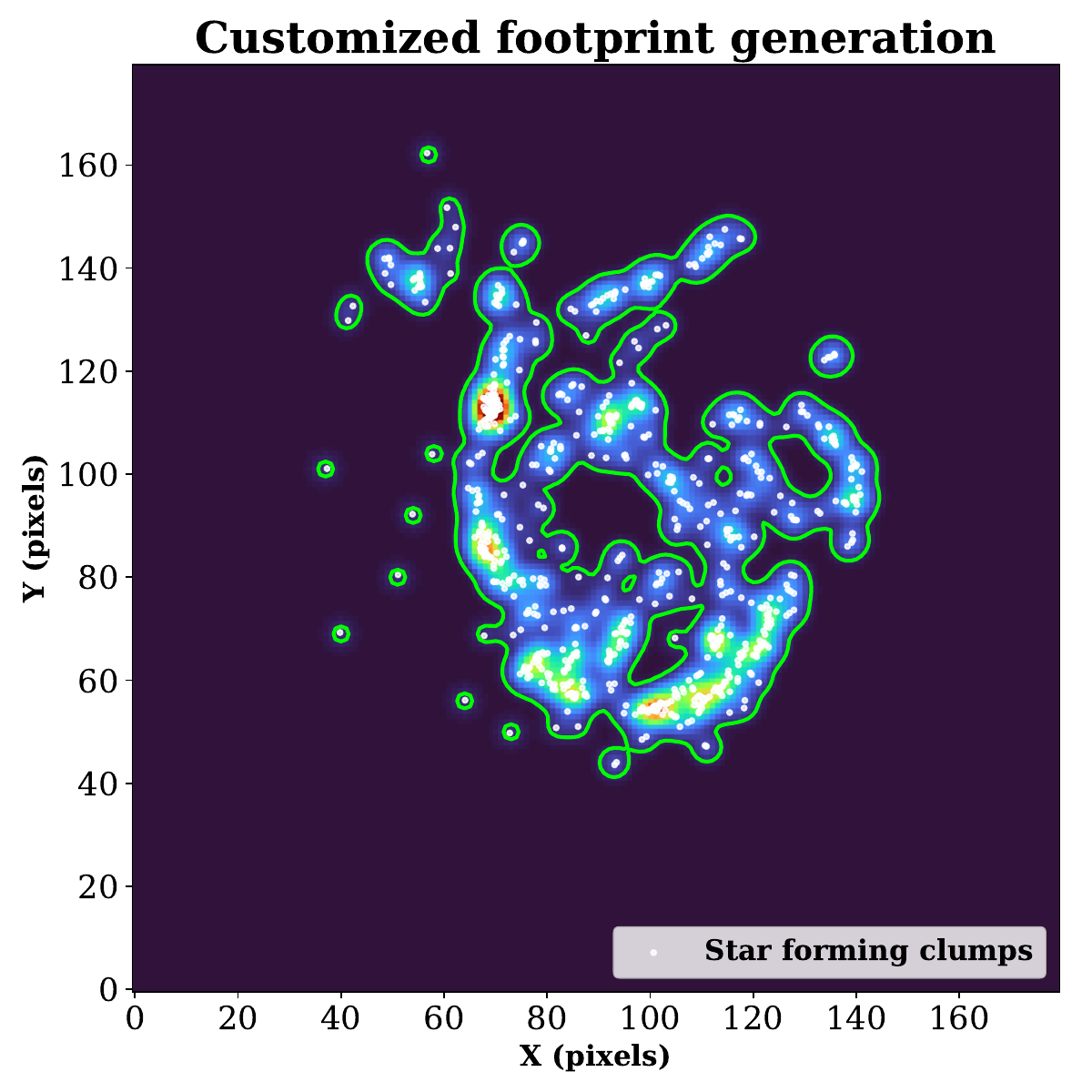}
    \caption{Schematic describing how the customised footprint for TPCF computation is generated using the positions of young SFCs in NGC 5457. The background image over which the contours are drawn is a 180$\times$180 pixel number density map of the SFC distribution. The white dots represent individual SFCs and the green contours mark the customised footprint of the SFC distribution within which random data points need to be populated for the TPCF calculation. See Section \ref{ss_TPCF}.6 for more details. }
    \label{footprint}
\end{figure}

For an accurate TPCF computation, it is recommended in \citet{2015ApJ...815...93G} that the random distribution exactly populates the same footprint as the data distribution (in our case, the de-projected positions of identified SFCs in the galaxy). To ensure this, we developed a novel method of generating a customised random distribution based on the unique distribution of SFCs within the galaxy. We created a pseudo fits image of the SFC distribution spanning 180$\times$180 pixels with 0 flux assigned to each pixel initially. Then, the SFC distribution was scaled down appropriately to fit within this 180$\times$180 pixel image. For every SFC located within a pixel, a flux value of 1 was added to that pixel and the pixels with no SFCs retained the flux value of 0. The resulting fits image was blurred using a Gaussian kernel of 2 pixel standard deviation. Next, flux surface density contours were drawn on this image, which serve as the customised footprint of the data distribution. Random points were then distributed within this customised footprint. Finally, this generated random distribution as well as the data distribution within the 180$\times$180 pixel image were appropriately scaled up to match the true spatial size of the galaxy and subsequently the TPCF computation could be performed. For illustration, the customised footprint generated from the distribution of young SFC positions in NGC 5457 is shown in Figure \ref{footprint}. The data as well as its corresponding random distribution are shown in Figure \ref{circular_v_custom}(left). We propose that our method of generating the customised random distribution sample is more accurate as compared to a regular random distribution (can be circular, elliptical, square etc). Our method yields efficient masking of low-density regions within the galaxy, reduces the underlying contribution of the galaxy disc and spiral arms in the TPCF computation and better reveals the true hierarchical distribution of the SFCs within the galaxy. The advantages of using a customised random distribution as opposed to a simple, circular random distribution in TPCF computation is discussed in Appendix \ref{apndx1}. \

Once the customised footprint was ready, we generated 100 random distributions within the footprint, all the while keeping the data points (de-projected positions of observed SFCs) fixed.  We ensured that each random distribution had approximately the same number of points as the data distribution. Thirty logarithmically spaced separation bins were taken for the TPCF calculation with the lower limit for the bins being twice the UVIT resolution and the upper limit being the largest separation between any two SFCs in the galaxy. We calculated the TPCF for all the 100 random distributions individually, as is given in equations 3, 4, and 5. For each spatial separation bin, the final TPCF and the associated error were taken to be the median and the standard deviation found in the TPCF calculated for the 100 random distributions, respectively.\\ 

\section{ Mathematical models for the observed two-point correlation function}
\label{ss_models}

\begin{figure*}
      \centering
	   \begin{subfigure}{0.42\linewidth}
		\includegraphics[width=\linewidth]{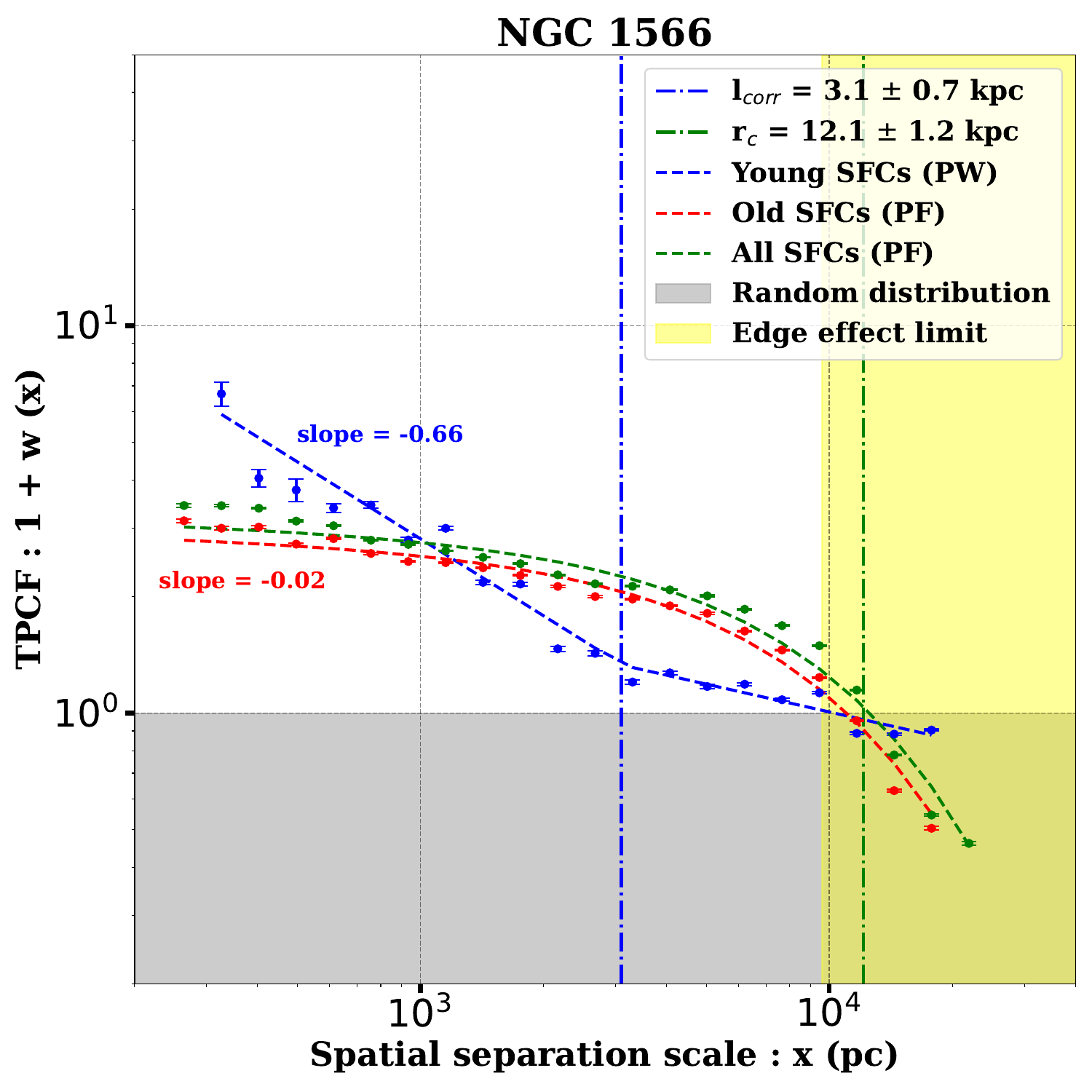}
		\label{fig:subfig1}
	   \end{subfigure}
	   \begin{subfigure}{0.42\linewidth}
		\includegraphics[width=\linewidth]{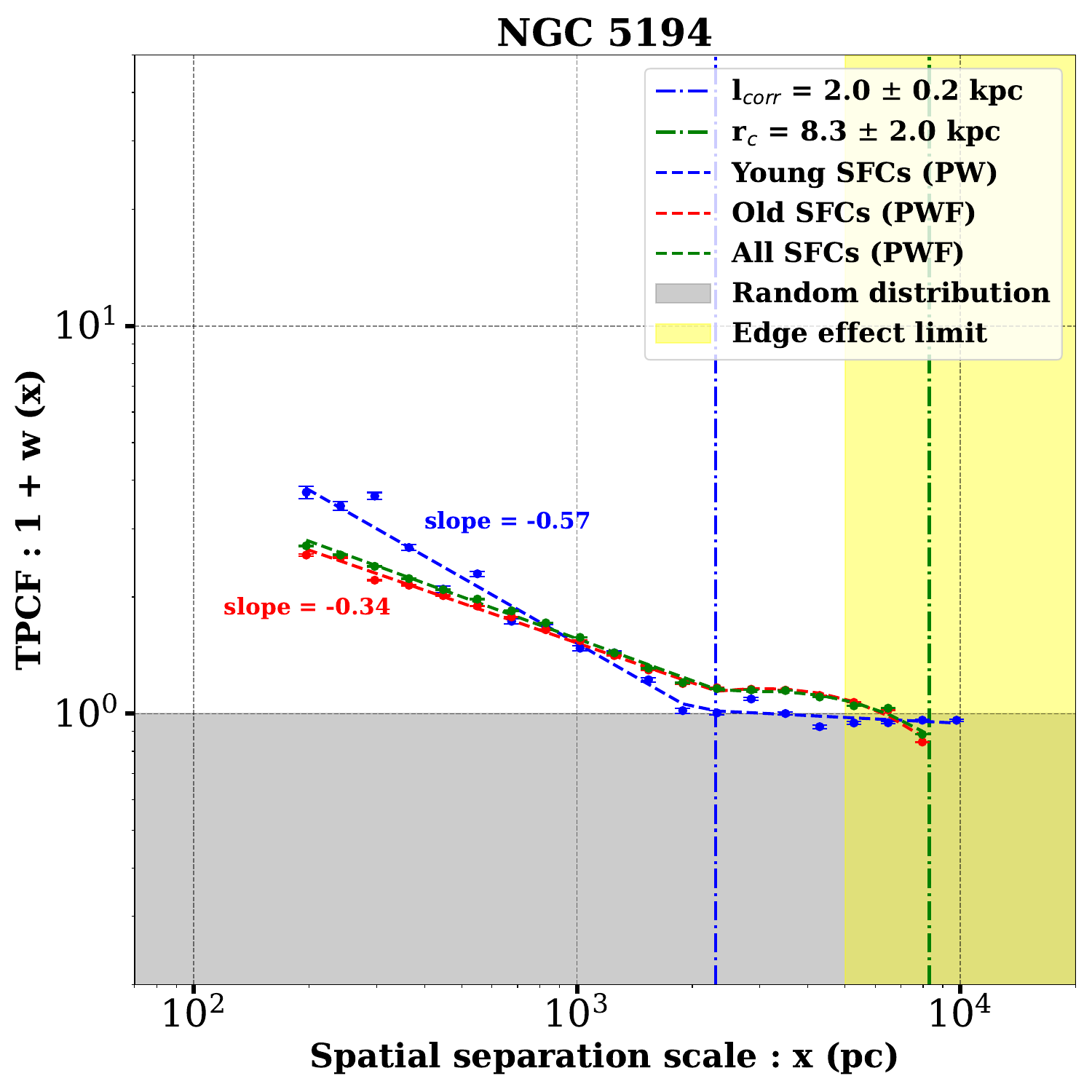}
		\label{fig:subfig2}
	    \end{subfigure}
	\vfill
	     \begin{subfigure}{0.42\linewidth}
		 \includegraphics[width=\linewidth]{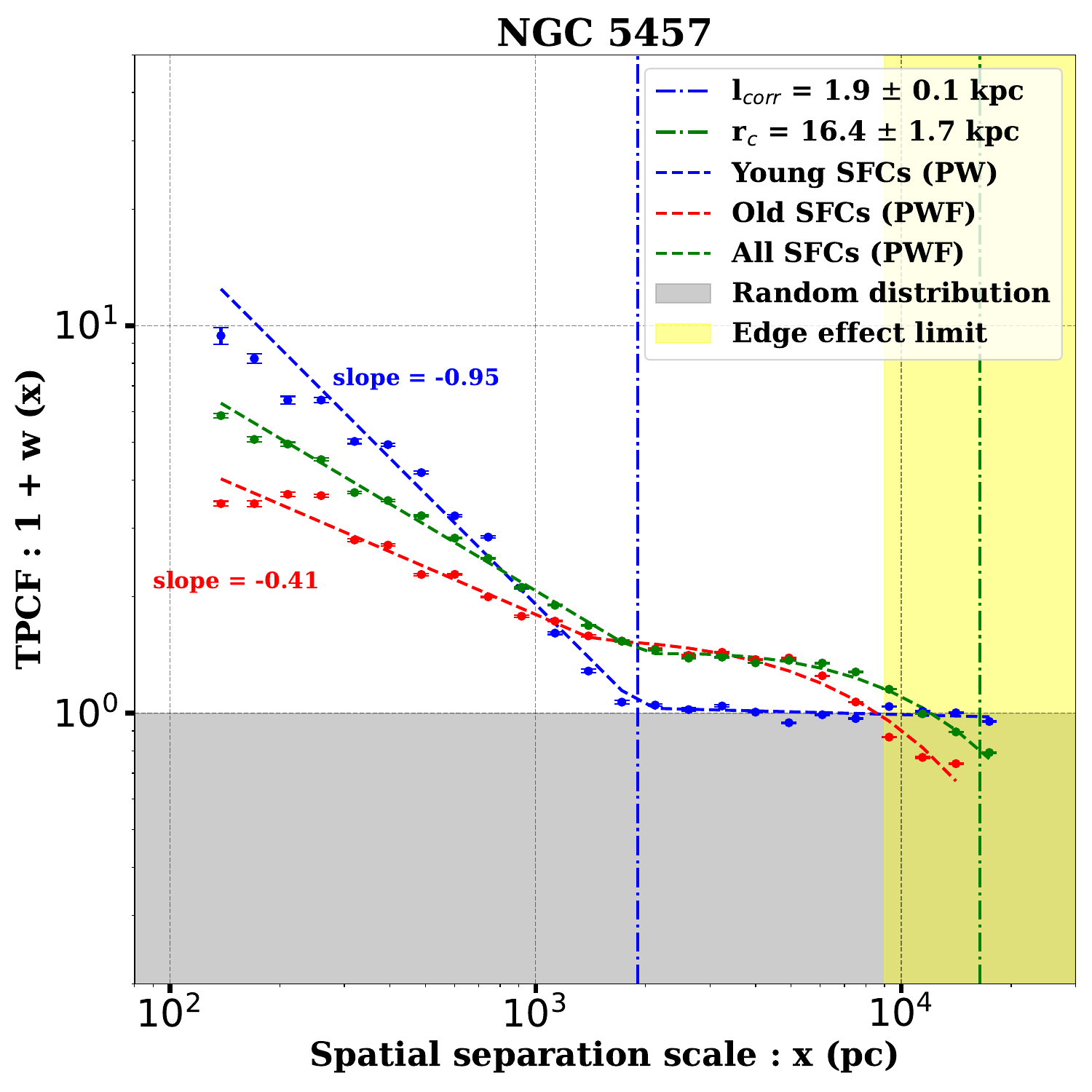}
		 \label{fig:subfig3}
	      \end{subfigure}
	       \begin{subfigure}{0.42\linewidth}
		  \includegraphics[width=\linewidth]{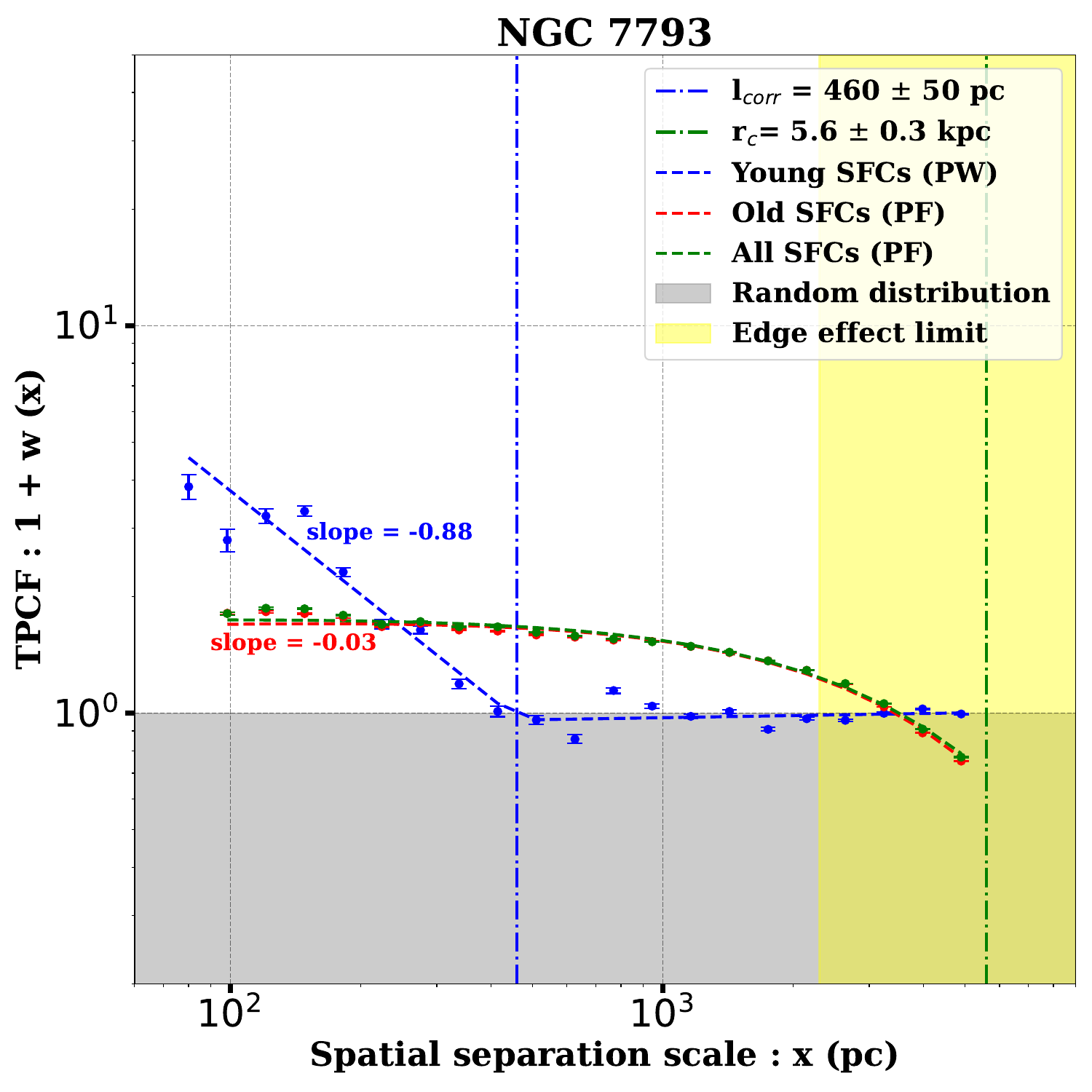}
		  \label{fig:subfig4}
	       \end{subfigure}
    \caption{Two-point correlation function (TPCF) as a function of spatial separation scale for the sample galaxies is shown in this figure for young (< 10 Myr), old (> 10 Myr) and all (young + old) SFCs. The mathematical models describing the observed TPCF plots (Section \ref{ss_models}) for a given population of SFCs are mentioned in the brackets. The hierarchy parameters such as $l_{\rm{corr}}$ and $r_c$ are also marked by dash-dotted blue and green lines. The slope of the TPCF plots for the young and old SFC distributions is quoted in blue and red colour, respectively. The shaded grey region has TPCF < 1 where the distribution of SFCs can be considered as Poissonian. The shaded yellow region represents the edge effect limit beyond which TPCF needs careful interpretation. This limit is equal to 1/5th of the largest separation between any two SFCs in the galaxy. The UVIT resolution limit for the galaxies is 25 - 130 pc and it is much smaller than the smallest separation seen in the TPCF plots of each galaxy.}
  \label{tpcf_plots}
\end{figure*}

We separately calculated TPCF for young, old, and all (young + old) SFCs and the emergent TPCF plots were found to be of the following three types (see Figure \ref{tpcf_plots}).
\\

1. Piecewise power law (Model PW) : This is a broken power-law model in which the first power law (valid on small scales) is quite steep but the second power law (valid on large scales, after the break) is usually very shallow (|slope| < 0.2). The break between the power laws is observed at a characteristic length scale, $l_{\rm{corr}}$. Physically, this model would correspond to a distribution that is clearly hierarchical on small scales but that shows signatures of being random on scales beyond $l_{\rm{corr}}$.
\begin{equation}
\label{eq:ModelPW}
\begin{array}{ll}
F_\mathrm{PW} (x) 
=\left\{\begin{array}{ll}
A_1 x^{\alpha_1} & :  x < l_{\rm{corr}}\\
A_2 x^{\alpha_2} & :  x > l_{\rm{corr}}
\end{array}\right.
\end{array}
.\end{equation}
Here, $A_1$ and $A_2$ are amplitudes related to each other by $A_2 = A_1 l_{\rm{corr}}^{(\alpha_1-\alpha_2)}$ and $\alpha_1$ and $\alpha_2$ are the two power-law slopes both of which are usually negative.\\

2. Power law with an exponential fall-off (Model PF): This is a single power-law model with quite a shallow negative slope, $\alpha_1$, and it undergoes an exponential fall-off at a characteristic length scale, $r_c$. As is shown by toy models in M21, this model can best describe a distribution of data points in a thin exponential disc that has a scale length approximately equal to $r_c$.
\begin{equation}
\label{eq:ModelPF}
\begin{array}{ll}
F_\mathrm{PF} (x) = A_1 x^{\alpha_1} \exp \left(- \frac{x}{r_c} \right)
\end{array}
.\end{equation}
Here, $A_1$ and $\alpha_1$ are the amplitude and slope of the power law and $r_c$ is the exponential fall-off scale.\\

3. Piecewise power law with an exponential fall-off (Model PWF): This TPCF model is a hybrid of model PW and model PF. In this model, we get a steep power law on smaller scales, which transitions to a shallow power law at $l_{\rm{corr}}$. The shallow power law in turn undergoes an exponential fall-off at $r_c$, where $r_c$ $\gg$ $l_{\rm{corr}}$. This model can be interpreted as arising from a distribution of data points that shows a strong hierarchical nature on small scales up to $l_{\rm{corr}}$, but is ultimately part of an exponential disc of scale length $r_c$.
\begin{equation}
\label{eq:ModelPW}
\begin{array}{ll}
F_\mathrm{PWF} (x) 
=\left\{\begin{array}{ll}
A_1 x^{\alpha_1} & : x < l_{\rm{corr}}\\
A_2 x^{\alpha_2} \exp \left(- \frac{x}{r_c} \right) & : x > l_{\rm{corr}}
\end{array}\right.
\end{array}
.\end{equation}
Here, $A_1$ and $A_2$ are two amplitudes related to each other by $A_2 = A_1 l_{\rm{corr}}^{(\alpha_1-\alpha_2)} \exp \left(+ \frac{l_{\rm{corr}}}{r_c} \right) $, $\alpha_1$ and $\alpha_2$ are the two negative power-law slopes, and $r_c$ is the exponential fall-off scale of the second power law.\\

The first two TPCF models -- namely, model PW and model PF -- are exactly similar to the M21 models of the same name (See section 4.2.1 of M21 for more details). The third TPCF model is a unique combination of model PW and model PF and we named it the PWF model. For the observed TPCF of a given population of SFCs, the model with the minimum reduced  $\chi^2$ value was used to explain it.\

We note that on spatial scales close to the instrument resolution, the number of correlated SFC pairs is found to be really small. The TPCF values corresponding to these scales were quite noisy and had error bars of the same order as the actual TPCF value. This can inhibit a robust computation of $l_{\rm{corr}}$ and $D_2$. Therefore, the first spatial separation bin considered in the fitting of observed TPCF was chosen to be the one after which the TPCF starts to fall monotonically as a function of the spatial separation scale. It typically corresponds to $\sim$3-4 times the UVIT resolution in all four galaxies (approximately 90 pc, 150 pc, 200 pc and 250 pc in NGC 7793, NGC 5457, NGC 5194, and NGC 1566, respectively). In the TPCF plots shown in this paper, we have masked the TPCF values for spatial scales smaller than the first spatial separation bin used in our analysis.\

\section{Results}
\label{s_results}

\textbf{\subsection{Qualitative interpretation of the observed two-point correlation function}}
\label{ss_interpretation}
The TPCF plots for young, old and all SFCs for the sample galaxies are shown in Figure \ref{tpcf_plots}. We found that in all four galaxies, Model PW best describes the TPCF of the young SFCs, so we fitted the observed TPCF with the PW model to compute $l_{\rm{corr}}$ and $D_2$ for each galaxy. The first power law corresponding to smaller scales usually has a steep negative slope, which is a clear signature of a purely hierarchical distribution. The second power law applicable to larger scales has approximately zero slope and it lies quite close to the absolute TPCF value of 1. This points to a nearly Poissonian distribution of star formation on large spatial scales. The break in the power law occurs at $l_{\rm{corr}}$ and it marks a clear physical transition from a hierarchical distribution to a nearly random one. Physically, the hierarchical distribution of star formation in a galaxy can be sustained up to $l_{\rm{corr}}$ by the various sources of ISM turbulence.\

For NGC 7793 and NGC 1566, the TPCF for old and all (young + old) SFCs were best described by the model PF. The TPCF plots for these populations of SFCs showed quite a small amount of absolute correlation, a shallow power law on smaller scales and an exponential fall-off on larger scales. The small absolute correlation can arise due to the distribution of SFCs within the galaxy disc and the exponential fall-off of the TPCF mimics the fall-off in the number of SFCs in the galaxy disc with increasing radial distance. On the other hand, in NGC 5194 and NGC 5457, the old and all SFCs showed model PWF type of TPCF, which implies that even some of the old SFCs in these galaxies are hierarchically distributed. We address this particular observation in more detail in Section \ref{ss_ageevolution}.4.\
\begin{figure*}
      \centering
	   \begin{subfigure}{0.43\linewidth}
		\includegraphics[width=\linewidth]{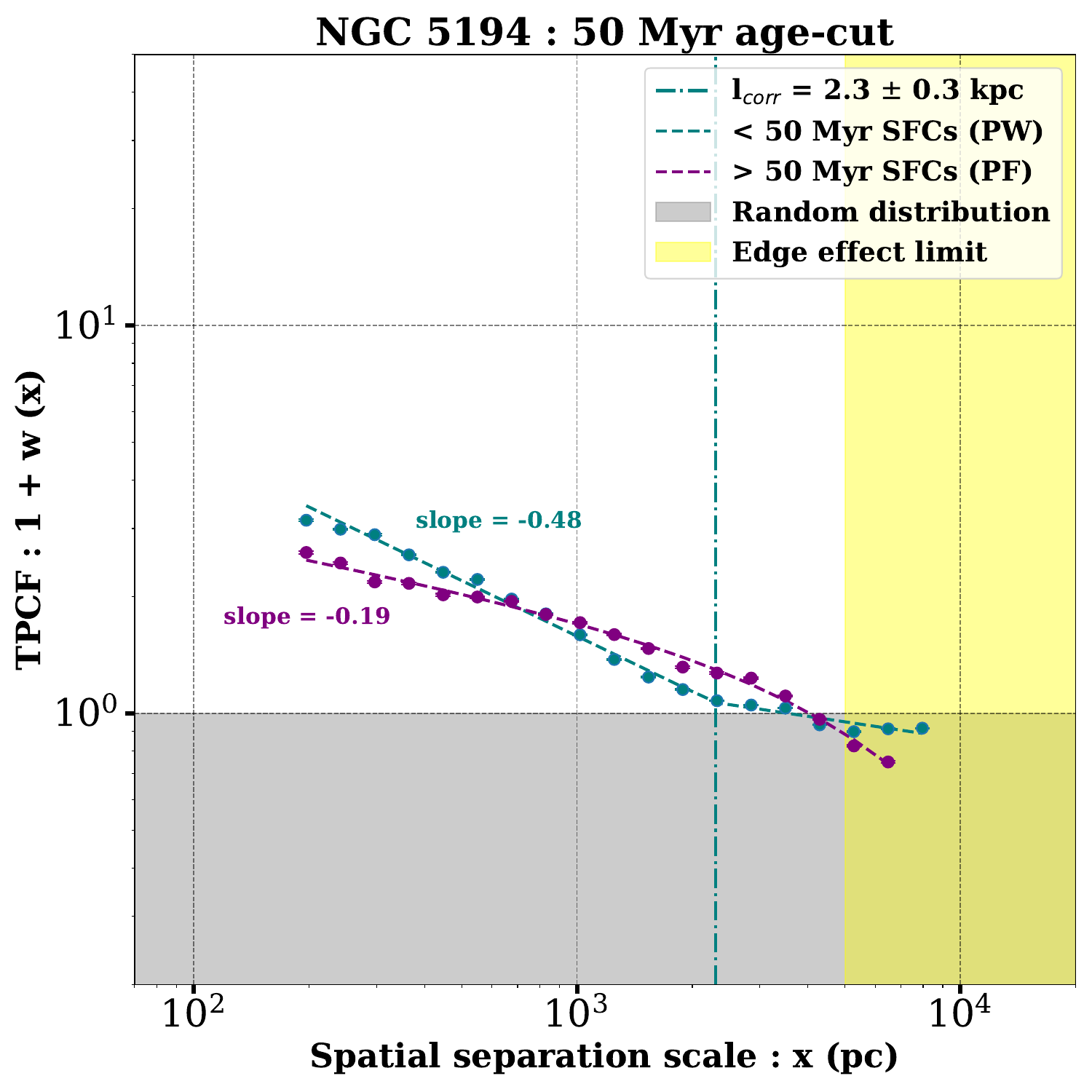}
		\label{fig:subfig1}
	   \end{subfigure}
	   \begin{subfigure}{0.43\linewidth}
		\includegraphics[width=\linewidth]{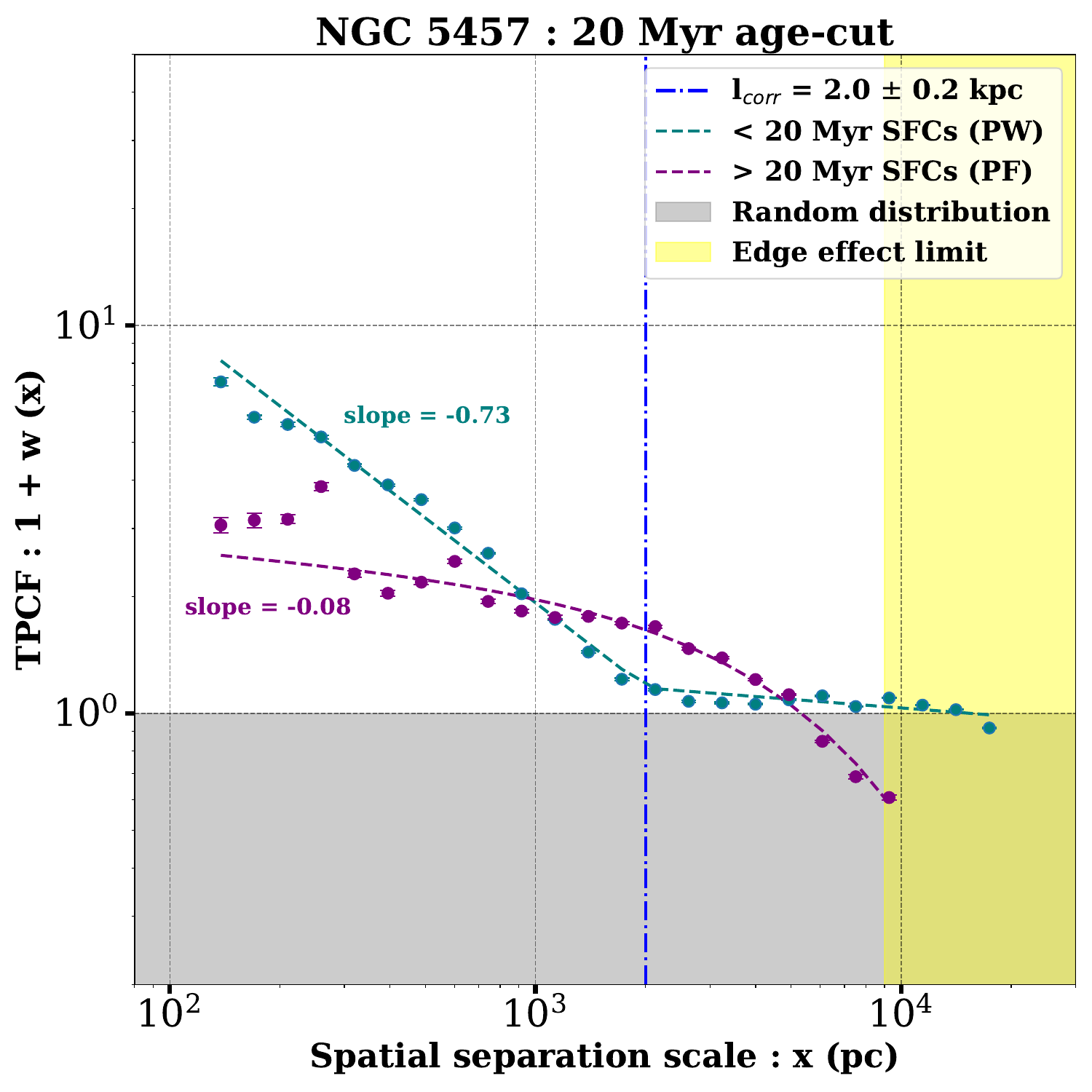}
		\label{fig:subfig2}
	    \end{subfigure}
    \caption{TPCF for NGC 5194 and NGC 5457 with a modified age-cut of 50 Myr and 20 Myr, respectively. We found that this new young population of SFCs shows hierarchical behaviour but the older population does not, as evidenced by the quite shallow slope of their TPCF. This infers that the hierarchical distribution of star formation dissipates in NGC 5194 and NGC 5457 roughly over timescales of 50 and 20 Myr, respectively. In addition, we found that even with the modified age-cuts, the $l_{\rm{corr}}$ and $D_2$ values for these galaxies do not change significantly in comparison to the values derived using the original 10 Myr age-cut to distinguish between young and old SFCs.}
    \label{tpcf_plots_mod}
\end{figure*}

In the observed TPCF plots, we have also shown the M21 edge-effect limit that is equal to one fifth of the largest separation between any two SFCs in the galaxy. Edge-effect limit is the length scale that arises due to the total spatial extent of the distribution. Beyond the edge-effect limit, TPCF needs to be interpreted with caution, particularly when a distribution of data points is truncated and is covered only partially. But, since we have full galaxy coverage with no truncation for the SFC distribution, the edge effect limit is not very significant in our results.\

\textbf{\subsection{Largest scale of hierarchy: $l_{\rm{corr}}$ }}
\label{ss_lcorr}
For the star formation hierarchy in a galaxy, $l_{\rm{corr}}$ demarcates the hierarchical regime of star formation on small scales from the stochastic regime of star formation on large scales. It is the largest scale of the star formation hierarchy and for our four sample galaxies, $l_{\rm{corr}}$ ranges from 0.5 kpc to 3.1 kpc (see Table \ref{table4}).\

For NGC 1566, we derived the $l_{\rm{corr}}$ value to be 3.1 $\pm$ 0.7 kpc, for which M21 could only compute the lower limit of $\sim$ 1.7 kpc. Similarly, for NGC 5194, our $l_{\rm{corr}}$ value of 2.0 $\pm$ 0.2 kpc is fairly consistent with the proposed lower limit of $\sim$ 2.7 kpc in M21. In NGC 7793, our $l_{\rm{corr}}$ value of 460 $\pm$ 50 pc is of the same order but slightly higher than the $\sim$100 pc value derived by M21. This can be due to a combined effect of several factors such as the different amount of galaxy coverage, a difference of star formation tracers and the relatively poorer resolution of the UVIT as compared to the HST. Lastly, NGC 5457 was only about 20\% covered in M21, so it is possible that the $l_{\rm{corr}}$ value for the entire galaxy can be quite different than the value computed for a significantly smaller region within the galaxy. In appendix \ref{apndx2}, we argue that the $l_{\rm{corr}}$ derived in M21 might not be the global value of $l_{\rm{corr}}$ in NGC 5457 rather, a local value. This can potentially explain the large difference in our $l_{\rm{corr}}$ value of 1.9 $\pm$ 0.1 kpc compared to the $\sim$ 450 pc value derived by M21. This result for NGC 5457 highlights the significant impact partial coverage of a galaxy can have on its hierarchy parameters.\

\textbf{\subsection{Fractal dimension of the hierarchy: $D_2$}}
\label{ss_D2}
We used the slope of the first power law ($\alpha_1$) of model PW for the young SFC TPCF to calculate the 2D fractal dimension of the star formation hierarchy ($D_2$ = 2 + $\alpha_1$) for our sample galaxies. The 3D fractal dimension of star-forming molecular clouds is often argued to be universal at 2.3 $\pm$ 0.3 (\citealt{1996ApJ...471..816E}; \citealt{Shadmehri_2011}). An idealised conversion of the fractal dimension from 3D to 2D ($D_3$ = 1 + $D_2$) suggests that the 2D fractal dimension of star formation should be close to $\sim$1.3. The $D_2$ values derived for our galaxies (Table \ref{table4}) range from 1.05 to 1.50, which represents quite a significant scatter. This implies that the fractal dimension of star formation hierarchies in different galaxies may be specific to the individual galaxy and might not be a universal value, unlike the fractal dimension of the molecular clouds. The significant variation that we found for $D_2$ is in line with the conclusions drawn by M21 as well as \citet{Sanchez_2008} and \citet{Grasha_2017_Spatial}. $D_2$ values computed for a much larger galaxy sample will be useful to further test the universality of the hierarchical properties of star formation in different galaxies.\

For NGC 1566, NGC 5194, and NGC 5457, our $D_2$ values agree well with the $D_2$ values from M21 but our values are found to be systematically smaller. This systematic decrease in $D_2$ is due to our improved, customised SFC distribution-based method of generating the random sample for the TPCF calculation and it is illustrated further in Appendix \ref{apndx1}. For NGC 7793, our first power law was found to be much shallower and consequently $D_2$ = 1.1 $\pm$ 0.1 is significantly larger than the M21 value of $D_2$ = 0.5 $\pm$ 0.3. Our shallower power law and larger $D_2$ can partly be explained by the fact that for NGC 7793, our $l_{\rm{corr}}$ value was also larger than the value derived by M21. The same factors of difference in star formation tracers, instrument resolution and galaxy coverage can contribute to the mismatch in $D_2$. But, as we have derived $D_2$ using the full galaxy coverage, our $D_2$ provides a better description of the star formation hierarchy in NGC 7793. The difference in the $D_2$ values calculated for NGC 1566, NGC 5194, and NGC 5457 using HII regions -- 1.77, 2.05, and 1.76, respectively (\citealt{Sanchez_2008}) -- in comparison to the $D_2$ values calculated using star clusters -- 1.5, 1.6, and 1.4 respectively (M21) -- and the $D_2$ values calculated using UVIT SFCs in this study -- 1.3, 1.4, and 1.1, respectively -- already highlights how different star formation tracers can lead to noticeable differences in the hierarchy parameters.\

\begin{table*}[t]
\caption{Derived hierarchy parameters using the TPCF models described in Section \ref{ss_models} which characterize the star formation hierarchy in the sample galaxies.}
\centering
\begin{tabular}{ccccccc}\hline
Galaxy &  Correlation length & Fractal dimension & Hierarchy dissipation & Modified $l_{\rm{corr}}$ & Modified $D_2$ & Exp. fall-off\\
& $l_{\rm{corr}}$ (kpc) &\newline ($D_2$ = 2 + $\alpha_1$) & timescale (Myr) & (kpc) & & scale : $r_c$  (kpc)  \\
(1) & (2) & (3) & (4) & (5) & (6) & (7) \\\hline
NGC 1566 & 3.1 $\pm$ 0.7 & 1.34 $\pm$ 0.08 & 10 & - & - & 12.1 $\pm$ 1.2 \\
NGC 5194 & 2.0 $\pm$ 0.2 & 1.43 $\pm$ 0.04 & 50 & 2.3 $\pm$ 0.3 & 1.52 $\pm$ 0.04 & 8.3 $\pm$ 2.0 \\
NGC 5457 & 1.9 $\pm$ 0.1 & 1.05 $\pm$ 0.04 & 20 & 2.0 $\pm$ 0.2 & 1.27 $\pm$ 0.04 & 16.4 $\pm$ 1.7 \\
NGC 7793 & 0.5 $\pm$ 0.1 & 1.12 $\pm$ 0.11 & 10 & - & - & 5.6 $\pm$ 0.3 \\\hline
\end{tabular}
\tablefoot{(1): Galaxy name. (2): $l_{\rm{corr}}$ represents the largest scale of the star formation hierarchy and it is calculated using the young (< 10 Myr) SFC TPCF. (3): $D_{2}$ is the fractal dimension of the hierarchy, which is calculated from the power-law slope $\alpha_1$ of the young SFC TPCF on scales smaller than $l_{\rm{corr}}$. (4): The hierarchy dissipation timescale is the age after which the SFCs have lost most of their initial, hierarchical distribution. (5) and (6): $l_{\rm{corr}}$ and $D_{2}$ values derived using the modified age-cut that is equal to the hierarchy dissipation timescale in column (4). For NGC 1566 and NGC 7793, the modified $l_{\rm{corr}}$ and $D_{2}$ values are same as the original $l_{\rm{corr}}$ and $D_{2}$, since the hierarchy dissipation timescale is 10 Myr for these galaxies. (7): $r_c$ is the exponential fall-off scale of the star formation distribution, calculated using all the SFCs (young + old) in the galaxy.}
\label{table4}
\end{table*}

\textbf{\subsection{Age evolution of the hierarchy and the hierarchy dissipation timescale}}
\label{ss_ageevolution}
M21 had illustrated for the star clusters of NGC 5194 that the power-law slope of the TPCF becomes increasingly shallower when progressively older star clusters are considered for TPCF calculation (Figure 3, M21). This indicated a gradual dissipation of the hierarchical distribution of the star clusters with age. In agreement with their results, we also found that the power-law slope of the TPCF declines with increasing age. Consequently, we expect that at sufficiently older ages, the slope of the TPCF approaches zero and the hierarchical distribution of SFCs within the galaxy gets completely dissipated. In our analysis, we assumed that a slope shallower than $-$0.2 represents a distribution that has lost most of its hierarchical properties.\

For NGC 5194 and NGC 5457, the old SFCs showed model PWF form of TPCF. The first power-law slope for old SFCs in these galaxies is significantly steeper than our assumed $-$0.2 slope cut of non-hierarchical distributions ($-$0.34 in NGC 5194 and $-$0.41 in NGC 5457). This implies that the older than 10 Myr population of SFCs has a non-negligible hierarchical distribution, which can suggest that the loss of the hierarchical structuring in these galaxies occurs over a timescale much larger than 10 Myr.\

To test this possibility, we used the fact that the hierarchy dissipation with age happens gradually. We assumed that at a certain age-cut greater than 10 Myr, say T Myr, the modified-old population (older than T Myr) will no longer show any hierarchical nature. In that case, the TPCF for this modified-old population could be fit with model PF instead of model PWF and the power-law slope for the corresponding model PF will be shallower than $-$0.2. To put this idea into practice, we gradually increased the age-cut for distinguishing between young versus old SFCs from 10 Myr, in steps of 10 Myr and re-computed the TPCF for the modified-old population each time. For the age-cut where this modified-old population TPCF can be best fit by the model PF with the power-law slope being shallower than $-$0.2, we concluded that the modified-old population has lost its initial hierarchical distribution. We found that the modified age-cut and thereby the hierarchy dissipation timescale for NGC 5194 and NGC 5457 is $\sim$50 Myr and $\sim$20 Myr, respectively. This can be seen in Figure \ref{tpcf_plots_mod}, in which the slope of the modified-old TPCF is shallower than $-$0.2. 

As is evident by the shallower than $-$0.2 slope of the old SFC TPCF plots in NGC 1566 and NGC 7793 in Figure \ref{tpcf_plots} as well as the TPCF plots for modified-old SFCs in NGC 5194 and NGC 5457 in Figure \ref{tpcf_plots_mod}, we infer that the hierarchy dissipation timescales for NGC 1566, NGC 5194, NGC 5457, and NGC 7793 are approximately 10, 50, 20, and 10 Myr, respectively. These timescales are also compiled in Table \ref{table4}. Incidentally, our 10 Myr and 50 Myr hierarchy dissipation timescales for NGC 7793 and NGC 5194, respectively, are in good agreement with the star cluster - molecular cloud disassociation timescale ($\sim$7 Myr for NGC 7793 (\citealt{Grasha_2018}) and $\sim$50 Myr for NGC 5194 (\citealt{Grasha_2019}) for these two galaxies. These authors argue that the shorter timescale in NGC 7793 compared to NGC 5194 could be due to its lower-pressure environment, which allows feedback from the star clusters to clear away the surrounding molecular gas on extremely short timescales. This agreement between our study and the aforementioned two studies shows the strong impact galaxy morphology and ambient environment can have on its star formation hierarchy and provides additional validity to our analysis. Our method of iteratively modifying the age-cut to find the transition of the modified-old SFC TPCF from model PWF (with power-law slope steeper than $-$0.2) to model PF (with power-law slope shallower than $-$0.2) is found to be very effective to determine the hierarchy dissipation timescale for a galaxy.\

Equipped with the hierarchy dissipation timescales for all the galaxies, we tested the effect of using 10 Myr age-cut versus the modified age-cut on the derived hierarchy parameters. We re-computed the hierarchy parameters $l_{\rm{corr}}$ and $D_2$ with the modified age-cuts of 50 Myr and 20 Myr in NGC 5194 and NGC 5457, respectively (Figure \ref{tpcf_plots_mod}). We found that even with the modified age-cut, our TPCF parameters do not change significantly. The TPCF slopes are found to be slightly shallower and consequently $D_2$ is slightly higher with the modified age-cut as compared to the original 10 Myr age-cut. This is consistent with the expectations of including older SFCs to the young ($<$ 10 Myr) SFC population. The modified $l_{\rm{corr}}$ values are found to show a marginal increase but stay within the error bars of the $l_{\rm{corr}}$ values derived using the original 10 Myr age-cut. This exercise indirectly verifies the robustness of our hierarchy parameters. For the sake of completeness, we have provided the $l_{\rm{corr}}$ and $D_2$ values derived with the modified age-cuts in Table \ref{table4}.\

We note that due to several assumptions associated with our age estimates using SB99, it is possible that some of the young SFCs have been misclassified as old in a few of our sample galaxies. This can also give rise to the hierarchical nature of the old SFC distribution in NGC 5194 and NGC 5457. But, the excellent match found between our hierarchy dissipation timescale with the star cluster - molecular cloud disassociation timescale for NGC 5194 and NGC 7793 indicates that the uncertainties associated with the SFC ages estimated using SB99 might not contribute to the variation in the observed TPCFs of our sample galaxies.\

\textbf{\subsection{Exponential fall-off scale of the star-forming clump distribution: $r_c$ }}
\label{ss_rc}
We found that for old and all (young + old) SFCs, on large separation scales, the TPCF falls off in an exponential tail with the fall-off scale being $r_c$. Since $r_c$ represents the disc distribution of the SFCs, we derived the $r_c$ values from the model PF or model PWF fit to the total population TPCF. We compared our $r_c$ values against the expectation that it should be comparable to the disc scale radius. We took the FUV disc scale radius of the galaxies NGC 5194, NGC 5457, and NGC 7793 from \citet{2017A&A...605A..18C} and found that our $r_c$ values are two to three times larger than the FUV scale radius. This suggests that the fall-off scale derived from the TPCF models might not exactly correspond to the disc scale radius, but a tight correlation between the two length scales can still be expected with a much bigger sample. The $r_c$ values derived for the sample galaxies is tabulated in Table \ref{table4}. We note that our old SFCs are not quite complete and therefore completeness might affect our estimates of $r_c$ of a galaxy.\\

\section{Discussion}
\label{s_discussion}
\textbf{\subsection{UVIT's view of hierarchical star formation}}
\label{HSFUVIT}
Through this study, we have demonstrated the feasibility of studying the hierarchical distribution of star formation in galaxies using UVIT, aided by its high resolution, sensitivity in FUV band and particularly large FoV. We achieved this by quantifying the nature of hierarchical star formation in four nearby spiral galaxies, NGC 1566, NGC 5194, NGC 5457, and NGC 7793. These four galaxies were previously studied in this context using the LEGUS-HST observations with partial galaxy coverage in M21. An accurate determination of the hierarchy parameters for these galaxies except NGC 7793 was hampered by the partial galaxy coverage available to M21. NGC 7793 itself was chosen in our study due to its near-complete galaxy coverage by the HST to act as a benchmark system for comparing UVIT and HST-based hierarchy parameters. The hierarchy parameters for NGC 7793 derived using the two instruments match reasonably well, which proves the robustness of our methodology.\ 

In this study, we have shown that the UV-bright, young SFCs are quite effective in probing the star formation hierarchy in a galaxy even though their average sizes are much greater than the sizes of the star clusters identified in the HST observations. We showed that UVIT data is an excellent probe for the investigation of hierarchical star formation by accurately determining the largest scale of the star formation hierarchy ($l_{\rm{corr}}$) for NGC 1566 and NGC 5194, which are found to be consistent with the lower limits found in M21. In M21 only lower limits could be given for $l_{\rm{corr}}$ and it was uncertain whether a largest scale of the hierarchy even exists for these galaxies. We have confirmed that every galaxy has a largest scale of the hierarchy and that it is much smaller than the size of the galaxy. It implies that there are indeed physical mechanisms acting within the galaxies that set the star formation hierarchies only up to $l_{\rm{corr}}$. To understand these exact physical mechanisms, a larger galaxy sample that looks at the correlation between $l_{\rm{corr}}$ and various large-scale galaxy properties is needed. We have also demonstrated in section \ref{ss_ageevolution}.4 that UVIT data can be effectively used to constrain the timescales over which the hierarchical distribution of the SFCs dissipates in different galaxies.\

\textbf{\subsection{$l_{\rm{corr}}$ in the context of previous studies}}
\label{comparison}

\begin{figure*}
    \centering
    \includegraphics[width=0.92\textwidth]{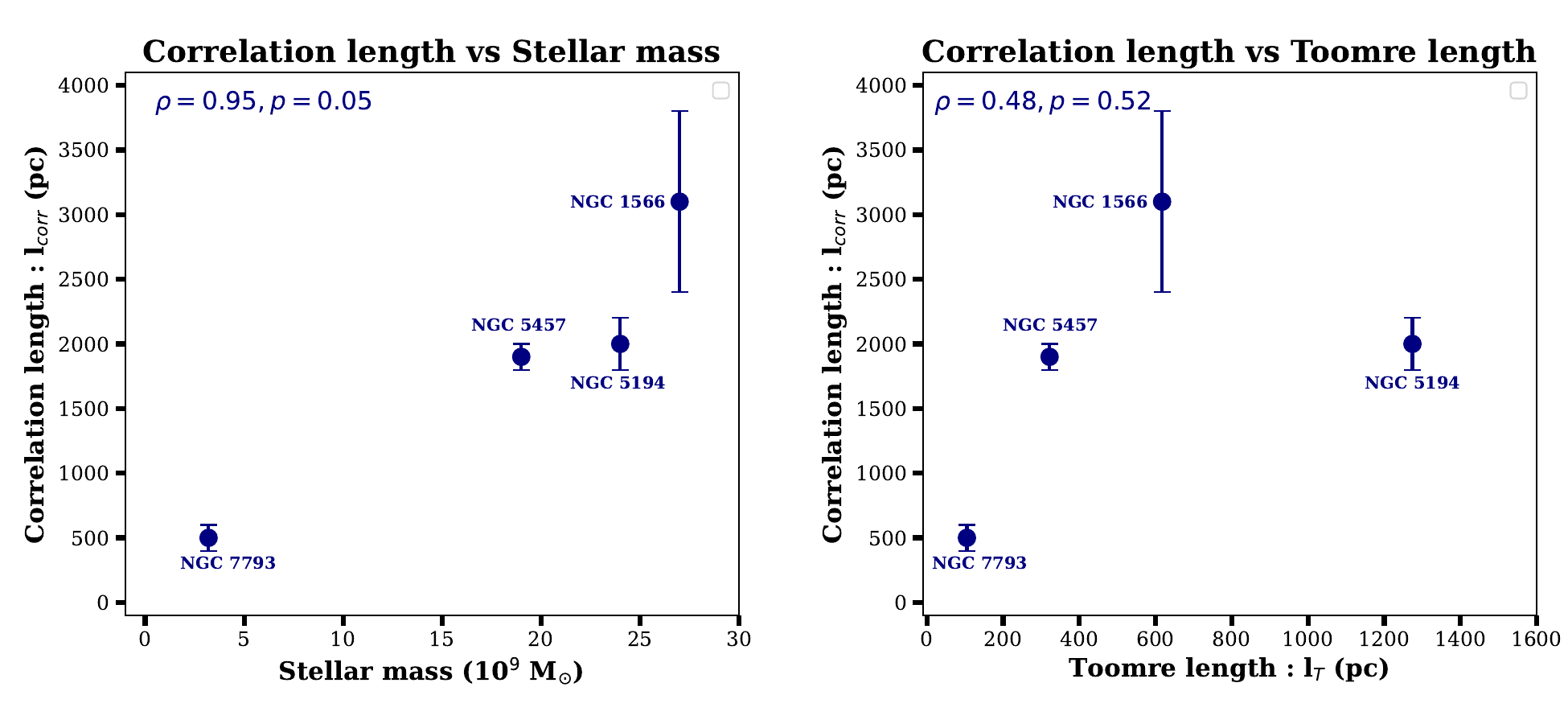}
    \caption{Trends of the estimated correlation length with the stellar mass (Left) and Toomre length (Right) of the sample galaxies. The Pearson correlation analysis indicates a strong positive correlation between the correlation length and the stellar mass of the galaxies, whereas only a mild positive correlation is observed between the correlation length and the Toomre length of the galaxies.
    }
    \label{Correlations_of_l_corr}
\end{figure*} 

With regard to why for NGC 1566 and NGC 5194 we have been able to estimate $l_{\rm{corr}}$ better than M21, to provide a consistent analysis across 12 sample galaxies, in M21, TPCF was computed only till 5 kpc. However, the considerably larger sizes of NGC 1566 and NGC 5194 and the partial galaxy coverage available in M21 can explain why these authors could only provide lower limits for $l_{\rm{corr}}$. In this paper, we were able to overcome these constraints by computing TPCF up to the spatial scales similar to the total galaxy size, which is only possible when the galaxy is fully covered in observations. UVIT's unique blend of high resolution and large FoV allowed us to probe a dynamic range of spatial separation scales on the hierarchical ($\lesssim$ 1 kpc) as well as the Poissonian ($\gtrsim$ 1 kpc) regime of the star formation hierarchy and to better identify the transition between the two regimes.\ 

We found that the $l_{\rm{corr}}$ value for NGC 7793 ($\sim$460 pc) is much smaller than those found for the other three galaxies. This could be because the shallower gravitational potential of NGC 7793 might not favour formation of really large, coherent star-forming structures, which is more probable in the deeper gravitational potential of the other three galaxies. This hypothesis agrees well with the positive correlation found between the galaxy stellar mass and $l_{\rm{corr}}$ in \citet{Grasha_2017_Spatial} and M21. We too observe for our four galaxies that the $l_{\rm{corr}}$ and the stellar mass of the galaxies follow a linear relationship (left panel of Figure \ref{Correlations_of_l_corr}). The Pearson correlation coefficient, $\rho$ = 0.95, and a $p$ value of 0.05 is indicative of a strong and statistically significant linear correlation between the $l_{\rm{corr}}$ and the stellar mass of the galaxies. A lack of strong spiral arms in NGC 7793 compared to the other three galaxies could also be a contributing factor in its lower $l_{\rm{corr}}$ value. The clustering properties, environmental conditions for star formation, and timescales of star formation are expected to be different with the nature of the spiral arms in the galaxy (\citealt{Grasha_2018}; \citealt{2024ApJ...964...13F}).\

Our $l_{\rm{corr}}$ values of $\sim$460 pc for NGC 7793 and $\sim$2.0 kpc for NGC 5194 compares favourably against the $R_{max}$ value from \citet{2017ApJ...842...25G} : $\sim$200 pc for NGC 7793 and $\sim$1kpc for NGC 5194, where $R_{max}$ is the length scale at which the age difference - separation ($\Delta t - R$) relationship flattens for the LEGUS star clusters. These authors suggested that $R_{max}$ is a rough estimate for the average size of the largest coherent star-forming regions physically possible in the star formation hierarchy of a galaxy. But what exactly determines ($l_{\rm{corr}}$) for a galaxy is still an open question. \citet{2017ApJ...842...25G} infer that $l_{\rm{corr}}$ can be related to the turbulent Jeans length of the galaxy, which can be set by large-scale self-gravity of the disc. Alternatively, $l_{\rm{corr}}$ can also be determined by some top-down mechanism such as galactic shear that can stop star-forming complexes from growing too large (\citealt{2008ApJ...685L..31E}; M21).\

In a differentially rotating, thin disc galaxy, shear due to galactic rotation can break overly large star-forming complexes into smaller ones and Toomre length is the length scale associated with this phenomenon (\citealt{1964ApJ...139.1217T}). For our sample galaxies, we compared the correlation length against the Toomre length (right panel of Figure \ref{Correlations_of_l_corr}), which is given by 
\begin{equation}
\label{eq:l_toom}
    l_\mathrm{T} =   \frac{4 \pi^{2} G \Sigma_\mathrm{g}} {\kappa^{2}} = \frac{4 \pi^{2} G \Sigma_\mathrm{g} r^{2}} {v^{2}}
.\end{equation}
Here, $G$ is the gravitational constant, $\Sigma_\mathrm{g}$ is the galaxy-averaged molecular hydrogen gas surface density, $\kappa$ is the epicyclic frequency, $r$ is the median galacto-centric radius of the SFCs, and $v$ is the flat rotation velocity of the galaxy. We directly took the values of $\Sigma_\mathrm{g}$ and $v$ used by M21 to estimate the Toomre length (see Appendix \ref{appdx4} for details). But our Toomre length values differ from those derived in M21 due to the choice of $r$, which was taken as the median galactocentric position of the SFCs in our analysis and the median galactocentric position of the star clusters in M21’s analysis. The complete galaxy coverage in our analysis compared to the partial galaxy coverage in M21 means that our $r$ values and consequently the Toomre length values are larger than those quoted in M21.\

The Toomre length values estimated here should represent the global star-forming conditions within our sample galaxies. Though the choice of $\Sigma_\mathrm{g}$, $r$ and $v$ can alter the absolute values of the Toomre length, we are only interested in the trends of $l_{\rm{corr}}$ with the Toomre length and not the absolute value of the Toomre length itself. We found only a mild positive correlation between the Toomre length and the correlation length of our sample galaxies (Pearson correlation coefficient, $\rho$ = 0.48, and a $p$ value of 0.52. This is in contrast with the results from M21, where a strong and statistically significant positive correlation between the two length scales was observed. However, it should be duly noted that the lack of a clear correlation can simply be a consequence of us having only four galaxies in our sample. Therefore, we aim to re-examine the relationship between the Toomre length and the correlation length with a larger, more diverse galaxy sample in future.\

\textbf{\subsection{Spatial variation in the hierarchical properties of star formation within galaxies}}
\label{spatialvariation}
For NGC 5457, we have estimated the global hierarchy parameters for the first time using UVIT observations. For NGC 5457, our $l_{\rm{corr}}$ value and the value derived in M21 are significantly different. We suggest that it is because their hierarchy parameters are only valid locally in the observed $\sim$20\% region of the galaxy and not valid for the full galaxy as a whole. We argue for this hypothesis by dividing NGC 5457 into four quadrants in Appendix \ref{apndx2} and computing $l_{\rm{corr}}$ for each quadrant. We found that the $l_{\rm{corr}}$ for the four quadrants are $\sim$1.2, 2.6, 3.1, and 1.8 kpc, respectively, which suggests that depending upon the star formation environment, the hierarchy parameters can vary significantly in different regions within the same galaxy. By dividing NGC 5194 into an inner and outer galaxy at a 4.7 kpc galacto-centric radius, \citet{Grasha_2019} found that the mass and median ages of the star clusters associated with the GMCs are significantly different in the inner and outer parts of the galaxy. In addition, they also found that the fractal dimension and correlation length in NGC 5194 for all the star clusters -- 1.60 and $\sim$2 kpc, respectively -- versus the same parameters for star clusters located within the smaller, Plateau de Bure Interefrometer Arcsecond Whirpool Survey (PAWS) footprint in the galaxy -- 1.72 and $\sim$200 pc, respectively -- were measurably different.\

\citet{2021ApJ...914...54Y} studied the star formation demographic in three extended-UV (XUV) disc galaxies including NGC 5457 using UVIT (though, they adopted a different method for SFC detection than us) and found that the SFCs in its outer disc were smaller and more compact than those present in the inner disc.  They invoke how global disc instabilities driving the star formation in the inner disc versus local disc instabilities driving the star formation in the outer disc could be leading to the observed differences. \citet{2024MNRAS.530.2199A} too found that in star-forming dwarf spirals, the outer SFCs have a smaller size, lower mass, and lower star formation rate density compared to the inner SFCs. The differences in the star formation properties with galacto-centric radius can arise due to higher gas density, mid-plane pressure, turbulence, and shear at small radii and vice versa. Similarly, azimuthal variations in the hierarchical properties of a galaxy can arise due to its interaction history and morphology. These results indicate that major differences between the local hierarchy parameters can exist within a single galaxy, which in turn might be different than the global hierarchy parameters of the galaxy. A more detailed investigation of the star formation hierarchies in different regions within galaxies for a bigger sample will be needed to explore this further.\

\textbf{\subsection{ Effect of completeness}}
\label{completenesseffects}
Incompleteness of the SFCs in the sample galaxies can potentially affect the derived hierarchy parameters. The evolution off the main sequence of progressively less massive OB stars present within an SFC with increasing age leads to a gradual fall in its FUV brightness. Deeper exposure images are complete up to fainter magnitudes, which would result in more number of fainter as well as older SFCs being detected. The completeness limits derived in Section \ref{ss_completeness}.4 suggest that most of our young ($<$10 Myr) SFC population with mass greater than $10^4 M_{\odot}$ is relatively complete. Since the hierarchy parameters, $l_{\rm{corr}}$ and $D_2$, are estimated based on these young SFCs, we expect incompleteness of the detected SFCs to have minimal impact on our final results.\\

\section{Summary}
\label{s_summary}
In this work, we have studied the nature of hierarchical star formation using the high angular resolution (1.5$^{\prime\prime}$) AstroSat UltraViolet Imaging Telescope (UVIT) FUV and NUV archival data of four nearby spiral galaxies, NGC 1566, NGC 5194, NGC 5457, and NGC 7793. The FUV observations from UVIT are highly sensitive in identifying the young star-forming regions in a galaxy, which in turn are excellent tracers of the hierarchical distribution of star formation within galaxies. We have shown that UVIT's distinct combination of a high resolution and a large 28$^\prime$ FoV makes it the ideal instrument with which we can simultaneously probe hierarchical star formation on small scales ($\sim$ 100 pc) as well as the large-scale ($\gg$ 1 kpc) distribution of star formation within the galaxy discs.\

We used the Astrodendro package to identify SFCs in the UVIT FUV images of the galaxies. Using Starburst99 simple stellar population models, we computed the ages of the identified SFCs and classified them as young (< 10 Myr) and old (> 10 Myr). By computing the spatial TPCF for these different populations of the SFCs, we have successfully derived the largest scale of the star formation hierarchy ($l_{\rm{corr}}$), the fractal dimension ($D_{2}$), and the hierarchy dissipation timescale for the sample galaxies and demonstrated the capabilities of UVIT in quantitatively studying hierarchical star formation. The main conclusions of our work are the following.\

    \begin{enumerate}
      \item We found that $l_{\rm{corr}}$, traced by the hierarchical distribution of young SFCs, ranges from $\sim$460 pc to $\sim$3.1 kpc in our sample galaxies. Our results suggest that the scale-free hierarchical distribution of star formation in galaxies has a largest scale: $l_{\rm{corr}}$, which is much smaller than the total galaxy size and is probably set by several physical mechanisms.\
      \item  In NGC 1566 and NGC 5194, we have derived the exact value of $l_{\rm{corr}}$ to be 3.1 and 2.0 kpc, respectively, instead of the lower limits found in \citet{2021MNRAS.507.5542M} (M21). In NGC 5457, we found significant variation in the global $l_{\rm{corr}}$ value ($\sim$1.9 kpc) and the local $l_{\rm{corr}}$ value ($\sim$450 pc) (derived in M21 for a small part of the galaxy). This highlights the need for full galaxy coverage in describing the hierarchical star formation within galaxies.\
      \item The $l_{\rm{corr}}$ value for the low-mass, flocculent spiral NGC 7793 is roughly five times smaller than the other three higher-mass, grand design spirals in our sample. We found a strong, statistically significant positive correlation of $l_{\rm{corr}}$ with the galaxy stellar mass and a mild positive correlation of $l_{\rm{corr}}$ with the Toomre length. These observations point towards the impact of the nature of spiral arms, the galaxy's gravitational potential, and shear due to galactic rotation in setting up the hierarchy of star formation.\
      \item The fractal dimension, $D_{2}$, of star formation in the sample galaxies spans a broad range from 1.05 to 1.50. 
      \item The fractal-like, hierarchical distribution of SFCs in our sample galaxies rapidly dissipates with age over timescales of 10$-$50 Myr.\
      \item The wide range of $l_{\rm{corr}}$, $D_{2}$, and hierarchy dissipation timescales found for our galaxies is indicative of non-universality in the hierarchical properties of star formation. This is contrary to the commonly assumed universality in the hierarchical properties of star-forming molecular clouds.\
   \end{enumerate} 

In our study, the FUV and NUV observations made using AstroSat-UVIT emerge as effective probes for studying star formation hierarchies, especially beyond kiloparsec scales. Anchoring future investigations on the UVIT FUV observations, this study can be extended to a much larger sample of star-forming galaxies to better understand the hierarchical process of star formation and its dynamic age evolution. In a direct follow-up to this study, we are investigating hierarchical star formation in approximately 30 galaxies spanning a large morphological range using archival UVIT data (Shashank et. al, in prep). We investigate the dependence of the hierarchy parameters on large-scale galaxy properties and examine whether the hierarchical properties of star formation are universal or dependent on the galaxy environment and morphology. This study will help us build a more comprehensive picture of the role that gravity, turbulence, galactic shear, galaxy dynamics, environment, and feedback play in the star formation cycle.\

\section*{DATA AVAILABILITY}

The AstroSat UVIT data used in this paper is publicly available at \url{https://astrobrowse.issdc.gov.in/astro_archive/archive/Home.jsp} and can be accessed with the proposal IDs G06\_087 for NGC 1566, A04\_176 for NGC 5194, 	G05\_233 for NGC 5457 and G06\_024 for NGC 7793. The science-ready UVIT fits images of the galaxies created in this study and the complete SFC catalog for each galaxy will be made public along with a forthcoming paper (Shashank et al, in prep) but can be shared in the meantime upon reasonable request to the corresponding author. 
\\

\begin{acknowledgements}
We thank the anonymous referee for a constructive report, which has helped us in improving our manuscript. GS thanks Renu, Chandan Watts, Rakshit Chauhan and Prajwel Joseph for their help during the methodology development stage. SS acknowledges support from the Science and Engineering Research Board
of India through the POWER research grant (SPG/2021/002672). S.H.M. acknowledge the support of NASA grant No. 80NSSC20K0500, NSF grant AST-2009679, and the Simons Foundation. C.M. acknowledges support from the National Science and Technology Council, Taiwan (grant NSTC 112-2112-M-001-027-MY3) and the Academia Sinica Investigator award (grant AS-IA-112-M04). This publication uses data from the UVIT, which is one of the key instruments on-board the AstroSat mission of the Indian Space Research Organisation (ISRO). The UVIT data is archived at the Indian Space Science Data Centre (ISSDC). We acknowledge the use of Python (\citealt{python09}), ASTROML (\citealt{2012cidu.conf...47V}), scikit-learn (\citealt{JMLR:v12:pedregosa11a}), Matplotlib (\citealt{Hunter07}), NumPy (\citealt{NumPy20}), SciPy (\citealt{SciPy20}), AstroPy (\citealt{astropy_2018}), Astrodendro (\href{http://www.dendrograms.org/}{http://www.dendrograms.org/)}, photutils (\citealt{larry_bradley_2024_10967176}) and CCDLAB (\citealt{2021JApA...42...30P}).
\end{acknowledgements} 

\bibliography{Paper}

\appendix
\onecolumn

\section{Circular versus Custom random sample}\label{apndx1}
\begin{figure}[ht!]
      \centering
	   \begin{subfigure}{0.33\linewidth}
		\includegraphics[width=\linewidth]{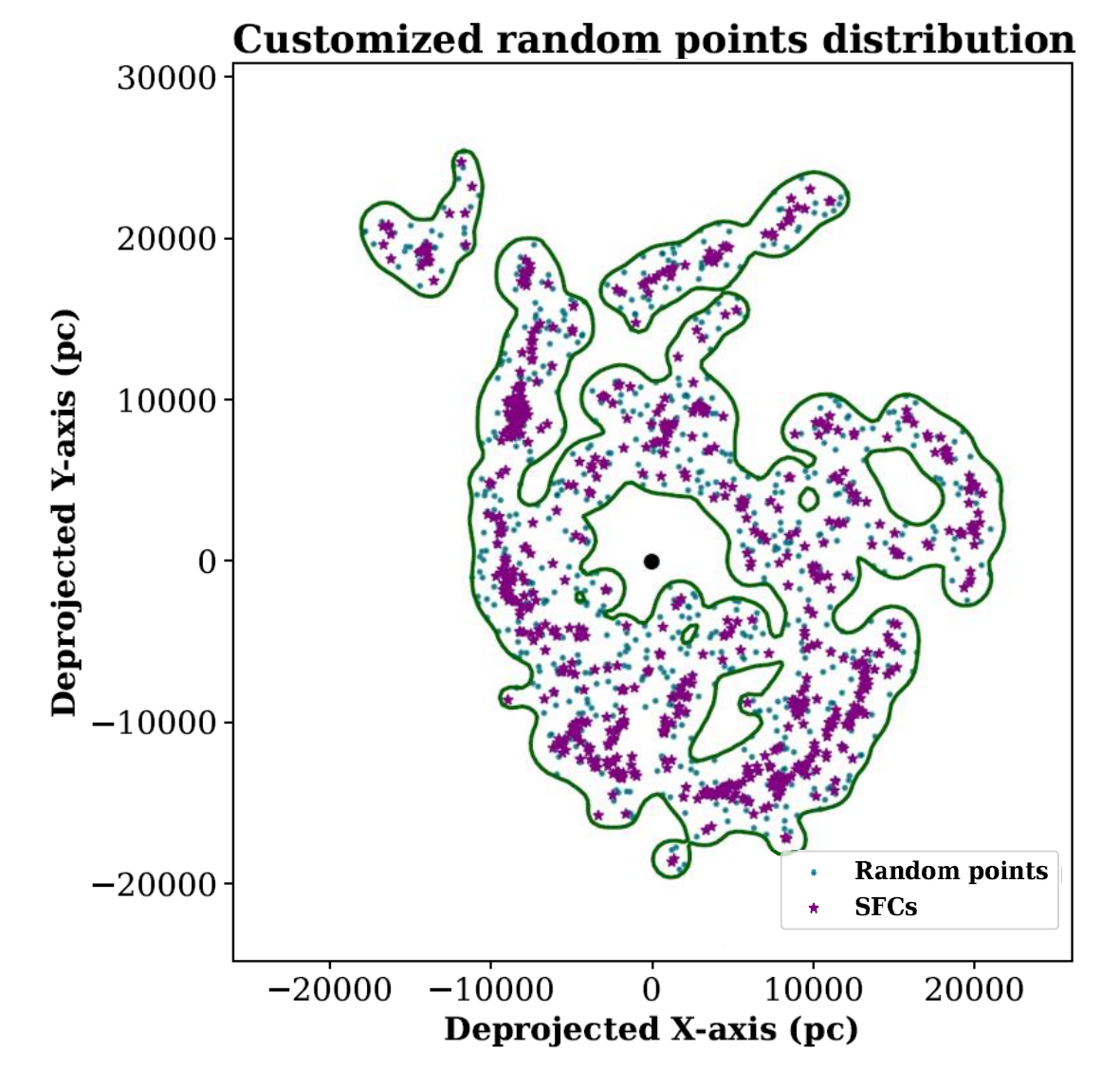}
		\label{fig:subfig1}
	   \end{subfigure}
	   \begin{subfigure}{0.33\linewidth}
		\includegraphics[width=\linewidth]{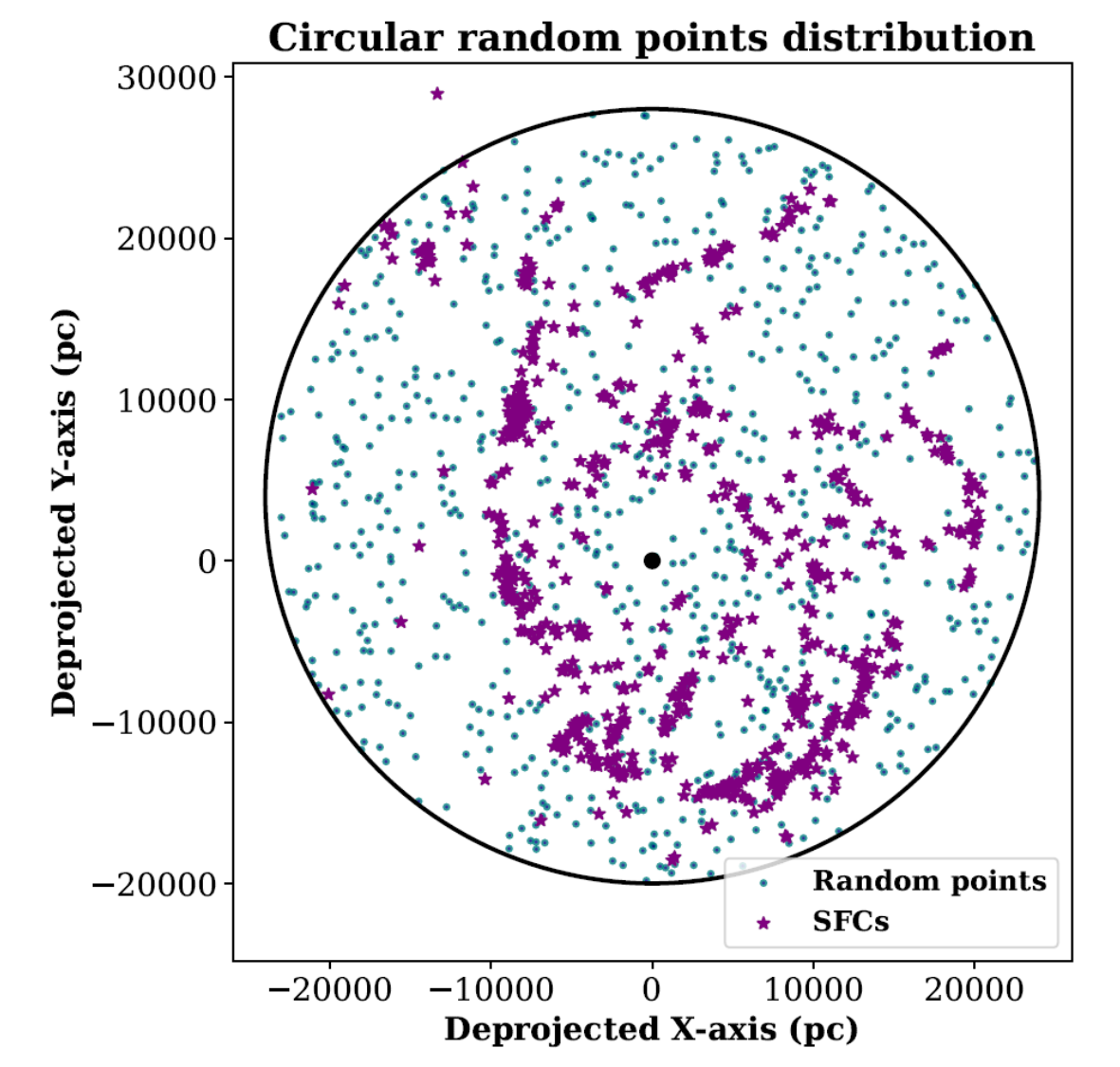}
		\label{fig:subfig2}
	   \end{subfigure}
  	   \begin{subfigure}{0.31\linewidth}
		\includegraphics[width=\linewidth]{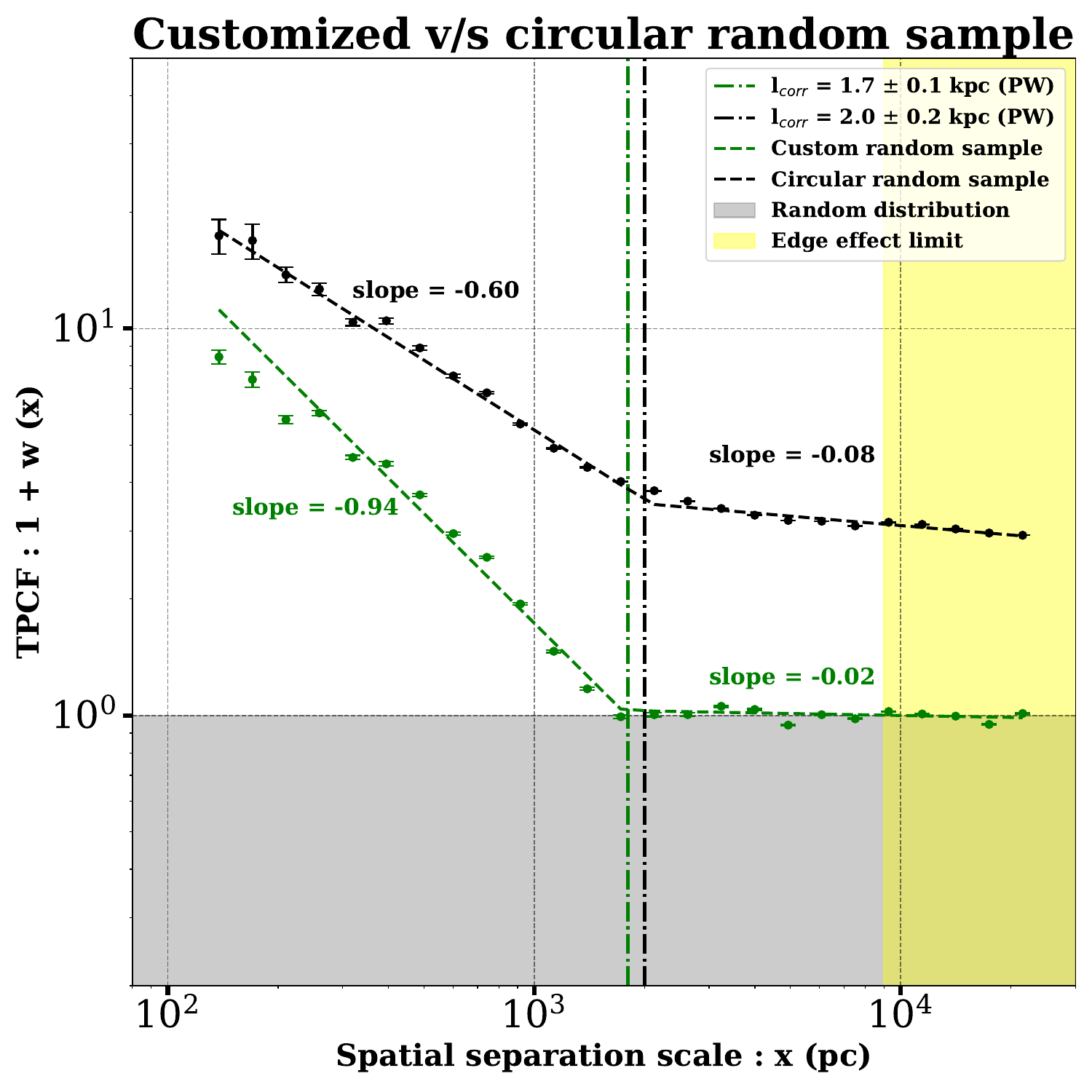}
		\label{fig:subfig2}
	    \end{subfigure}
    \caption{Effect of a customised versus circular random sample on the TPCF of young SFCs in NGC 5457: Even though the $l_{\rm{corr}}$ value is found to be nearly the same in both the cases, the first power-law slope is found to be shallower for a circular random sample. The absolute correlation and the slope of the second power law are found to be considerably higher in case of circular random sample. In the left panel, green contours mark the customised footprint within which random points are populated. In the left and central panel, purple stars and cyan dots represent the true SFC positions and the random points generated for TPCF calculation, respectively.}
    \label{circular_v_custom}
\end{figure}

To reliably compute the TPCF of a distribution of data points, it is crucial that the random distribution exactly populates the footprint of the actual data. We studied the impact of choosing a regular (circular in this case) random distribution versus choosing a customised random distribution on the TPCF of the young SFC population in NGC 5457. As can be seen in Figure \ref{circular_v_custom} (middle panel), the circular random distribution does not exactly follow the data distribution. For the same data distribution, we created a customised random distribution (Figure \ref{circular_v_custom}; left panel) by the method explained in section \ref{ss_TPCF}.6. The resulting TPCF using the two kinds of random distributions is also shown in Figure \ref{circular_v_custom} (right panel). The $l_{\rm{corr}}$ value is found to be quite similar for both the random distributions. The circular random distribution leads to a shallower first power law. This results in a relatively smaller $D_2$ for the customised random distribution, which is what we found when comparing our $D_2$ for NGC 1566, NGC 5194 and NGC 5457 with the M21 $D_2$ values. For the circular random distribution, the second power law has a non-zero slope indicative of a residual hierarchical nature. But when a customised random distribution is used, the second power law has a nearly zero slope and the data points follow a non-hierarchical distribution on these scales. We propose that our method of generating the customised random distribution can better remove the contribution of spiral arms and exponential disc from the TPCF calculation. The choice of a regular random distribution (circular in our illustration, but can be elliptical, square etc.) can explain why in M21, the second power law for many galaxies had a non-zero slope and had more absolute correlation than expected from a completely Poissonian distribution on larger scales. Our method of random distribution generation can be very effective in future for galaxies with irregular or distorted morphologies.\ 

\section{Local versus Global hierarchical parameters}
\label{apndx2}
\begin{figure}[ht!]
      \centering
	   \begin{subfigure}{0.48\linewidth}
		\includegraphics[width=\linewidth]{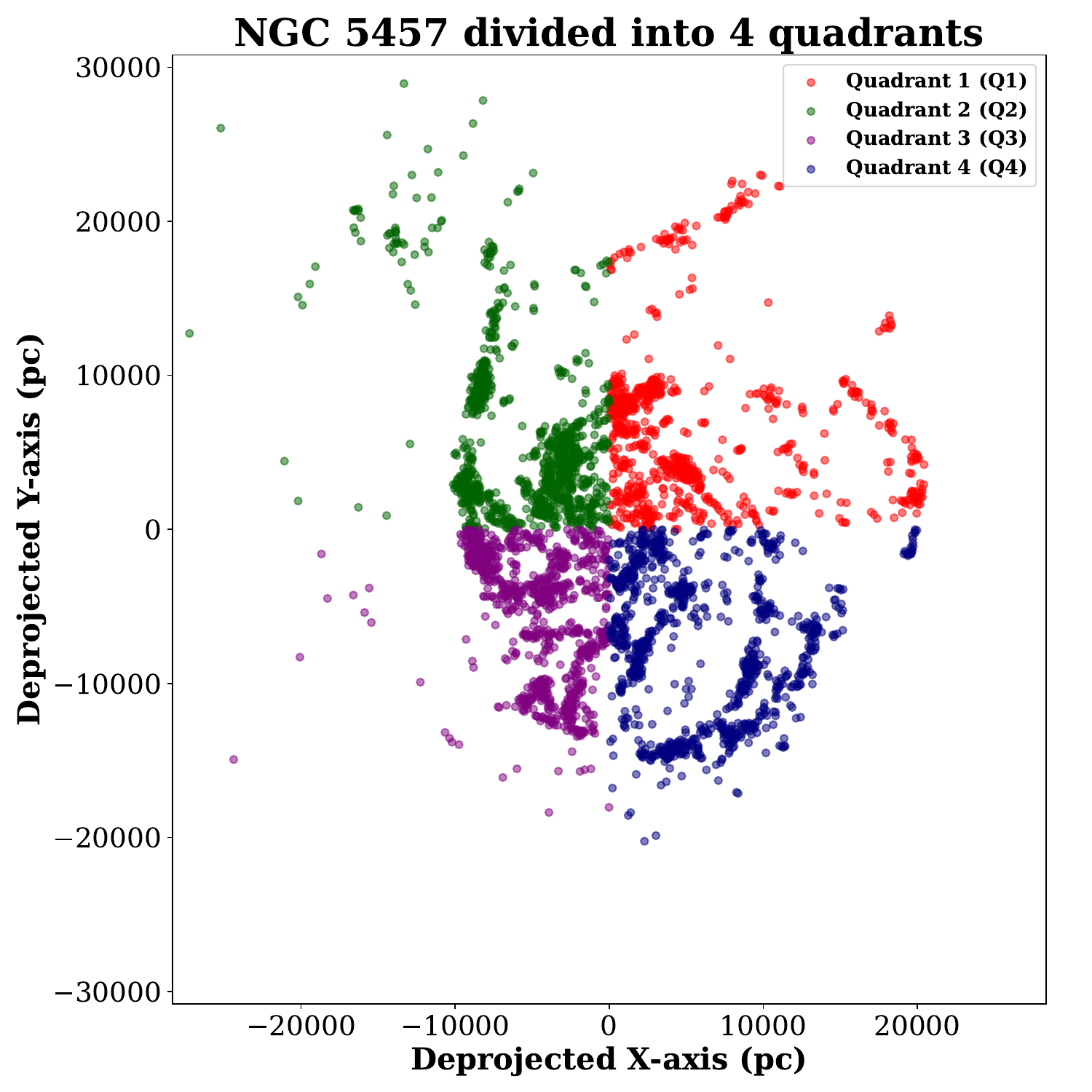}
		\label{fig:subfig1}
	   \end{subfigure}
	   \begin{subfigure}{0.48\linewidth}
		\includegraphics[width=\linewidth]{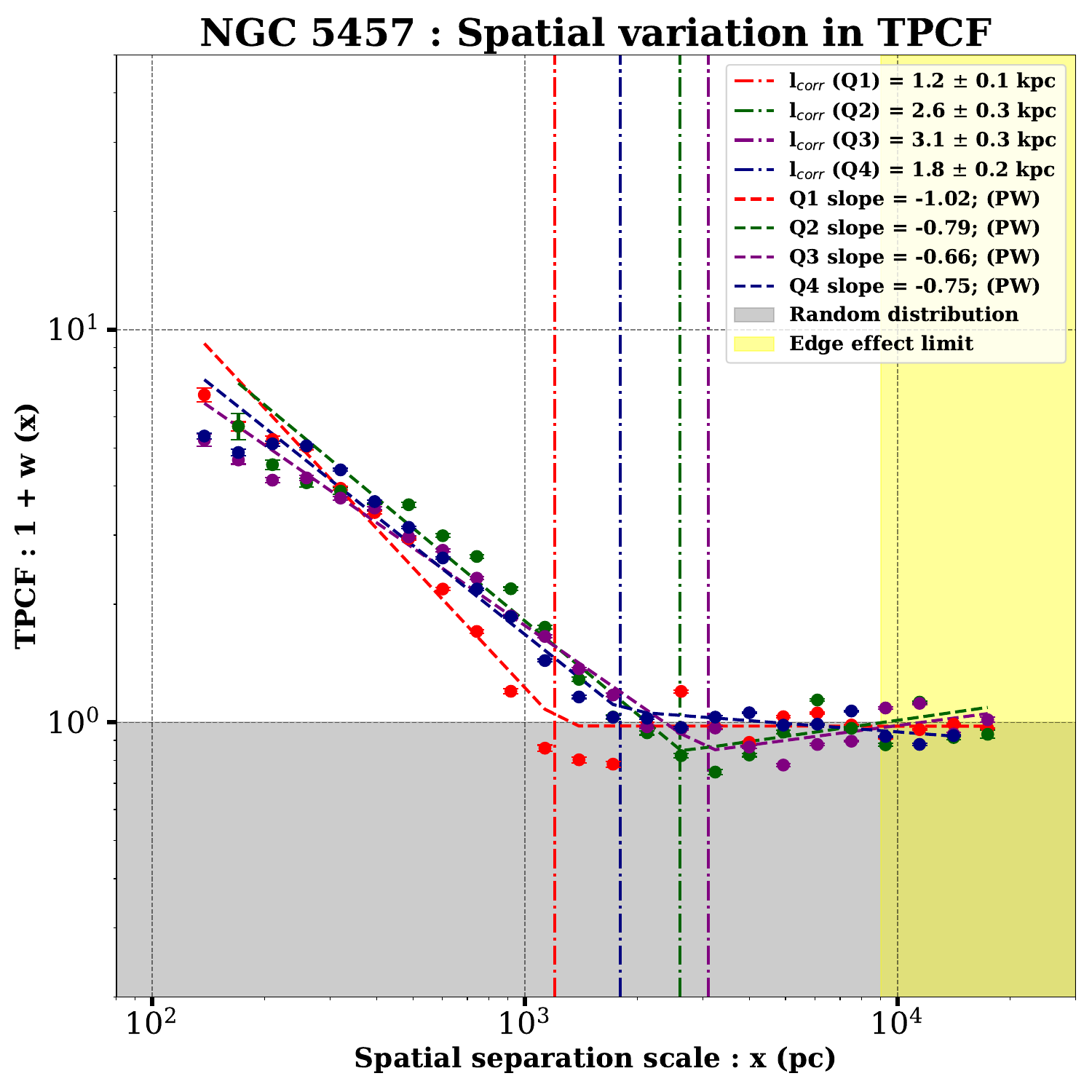}
		\label{fig:subfig2}
	    \end{subfigure}
    \caption{Spatial variation in the TPCF of NGC 5457 (SFCs with magnitude error < 0.15): The four quadrants of NGC 5457 have quite different values of $l_{\rm{corr}}$ from each other. This implies that the conditions of star formation can be significantly different in different parts of the galaxy and the global hierarchical parameters of a galaxy can be quite different than the local hierarchy parameters.}
    \label{global_v_local}
\end{figure}

NGC 5457 had only a small fraction ($\sim$20\%) of the galaxy area covered in the HST observations of M21 but UVIT provides us the full galaxy coverage. We found that the $l_{\rm{corr}}$ value for NGC 5457 using the two datasets are quite different (450 pc by HST, 1.9 kpc by UVIT). This discrepancy could be mainly due to the global hierarchy parameters of any galaxy being quite different from the local hierarchy parameters computed from a small part of the galaxy. This could not be tested directly because, in the HST-covered region of NGC 5457, we did not have enough young UVIT SFCs to enable a robust computation of TPCF and calculation of $l_{\rm{corr}}$. Therefore, to address the discrepancy between global and local hierarchy parameters, we divided the de-projected positions of 0.15 magnitude error cut SFCs in NGC 5457 into four quadrants as seen in Figure \ref{global_v_local} (left panel). We computed $l_{\rm{corr}}$ from the young SFCs present in each quadrant. We found that the $l_{\rm{corr}}$ with 0.15 magnitude error cut is $\sim$ 1.2, 2.6, 3.1 and 1.8 kpc for quadrants 1, 2, 3 and 4 (Figure \ref{global_v_local} : right panel). The $l_{\rm{corr}}$ with 0.15 magnitude error cut for all the young SFCs (all quadrants combined) is $\sim$ 2.7 kpc (not shown here). These five different values of $l_{\rm{corr}}$ represent a significant variation in the properties of the star formation hierarchy within a single galaxy. Our analysis suggests that the star formation conditions and therefore $l_{\rm{corr}}$ can be significantly different in different parts of the galaxy. Our quadrant-based analysis indirectly proves that the hierarchy parameters are prone to change if the galaxy is not fully covered in the observations. This further emphasises the importance of having full galaxy coverage in computing the hierarchy parameters of a galaxy and thereby the usefulness of an instrument like UVIT with a large FoV. \ 

We note in passing that for NGC 5194, where a sufficient number of young UVIT SFCs are present in the HST-covered region of the galaxy, our $l_{\rm{corr}}$ value of $\sim$1.4 kpc in the HST-covered region is smaller than the 2.0 kpc value calculated for the whole galaxy using UVIT. This is again suggestive of the variation between global and local hierarchical properties of a galaxy.\

\section{Synthetic versus observed colour-magnitude diagrams for NGC 1566, NGC 5194, and NGC 5457}
\label{appdx3}
The synthetic CMDs generated using SB99 and the CMD of the observed SFCs in NGC 1566, NGC 5194 and NGC 5457 are presented in this section. Unlike NGC 7793, all of these three galaxies have similar input parameters in SB99 including the metallicity Z = 0.02. Therefore, the colours and evolutionary tracks are observed to be the same for all three galaxies. However, the distance-dependent synthetic FUV magnitudes in the evolutionary tracks, extinction-corrected observed magnitudes and the mass range of the observed SFCs are found to be varying amongst galaxies, as was expected.\

\begin{figure}[ht!]
      \centering
	   \begin{subfigure}{0.33\linewidth}
		\includegraphics[width=\linewidth]{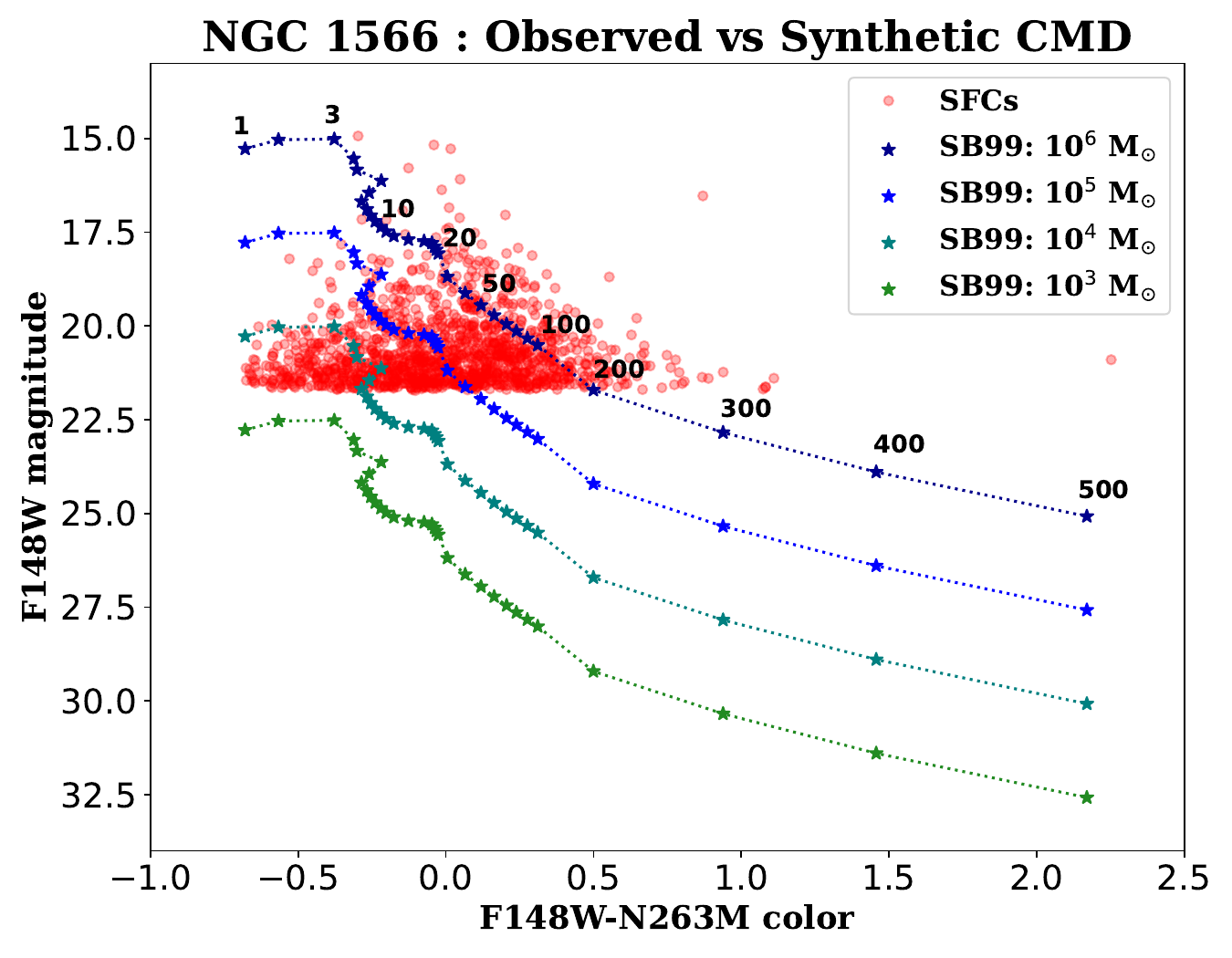}
		\label{fig:subfig1}
	   \end{subfigure}
	   \begin{subfigure}{0.33\linewidth}
		\includegraphics[width=\linewidth]{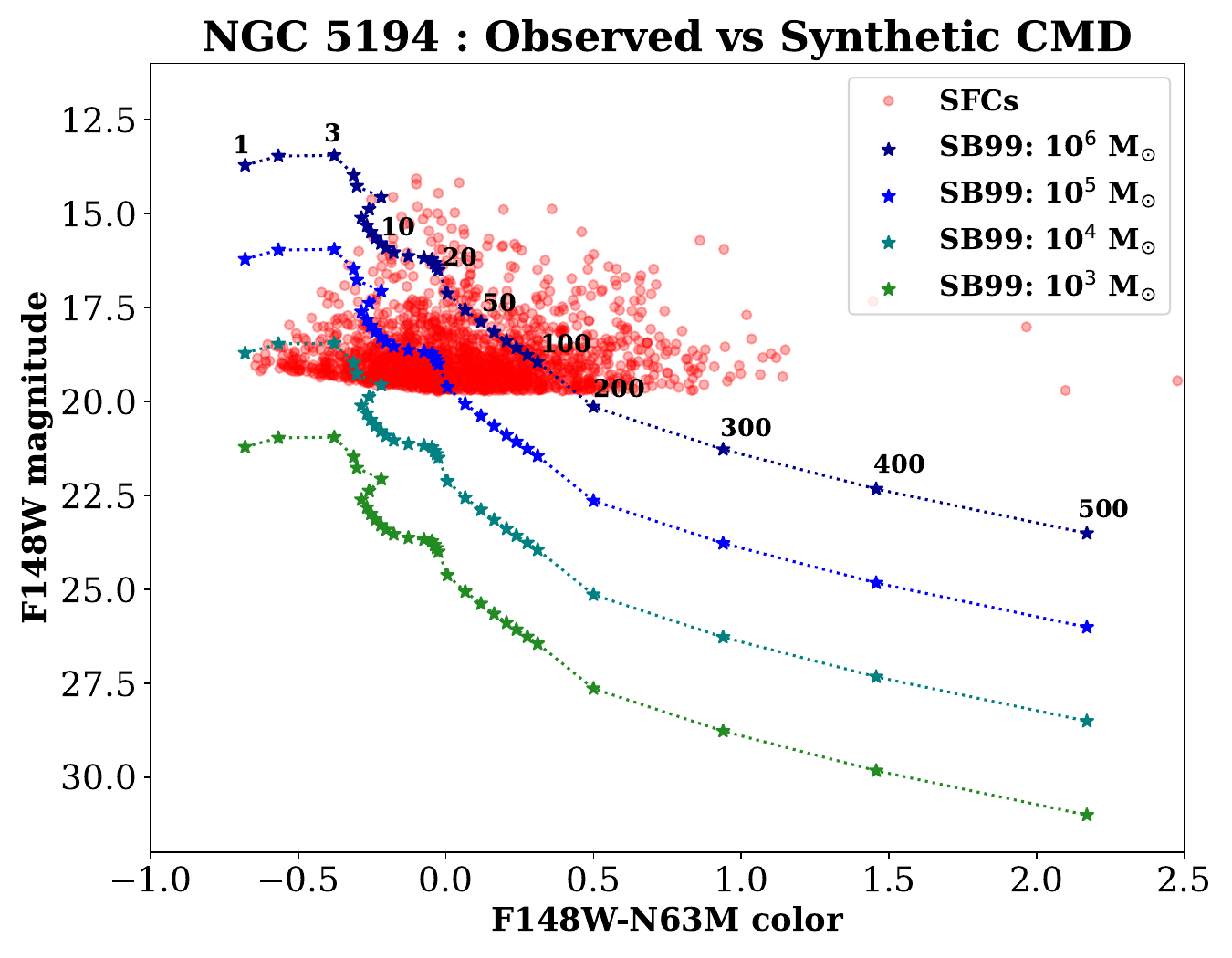}
		\label{fig:subfig2}
	   \end{subfigure}
  	   \begin{subfigure}{0.33\linewidth}
		\includegraphics[width=\linewidth]{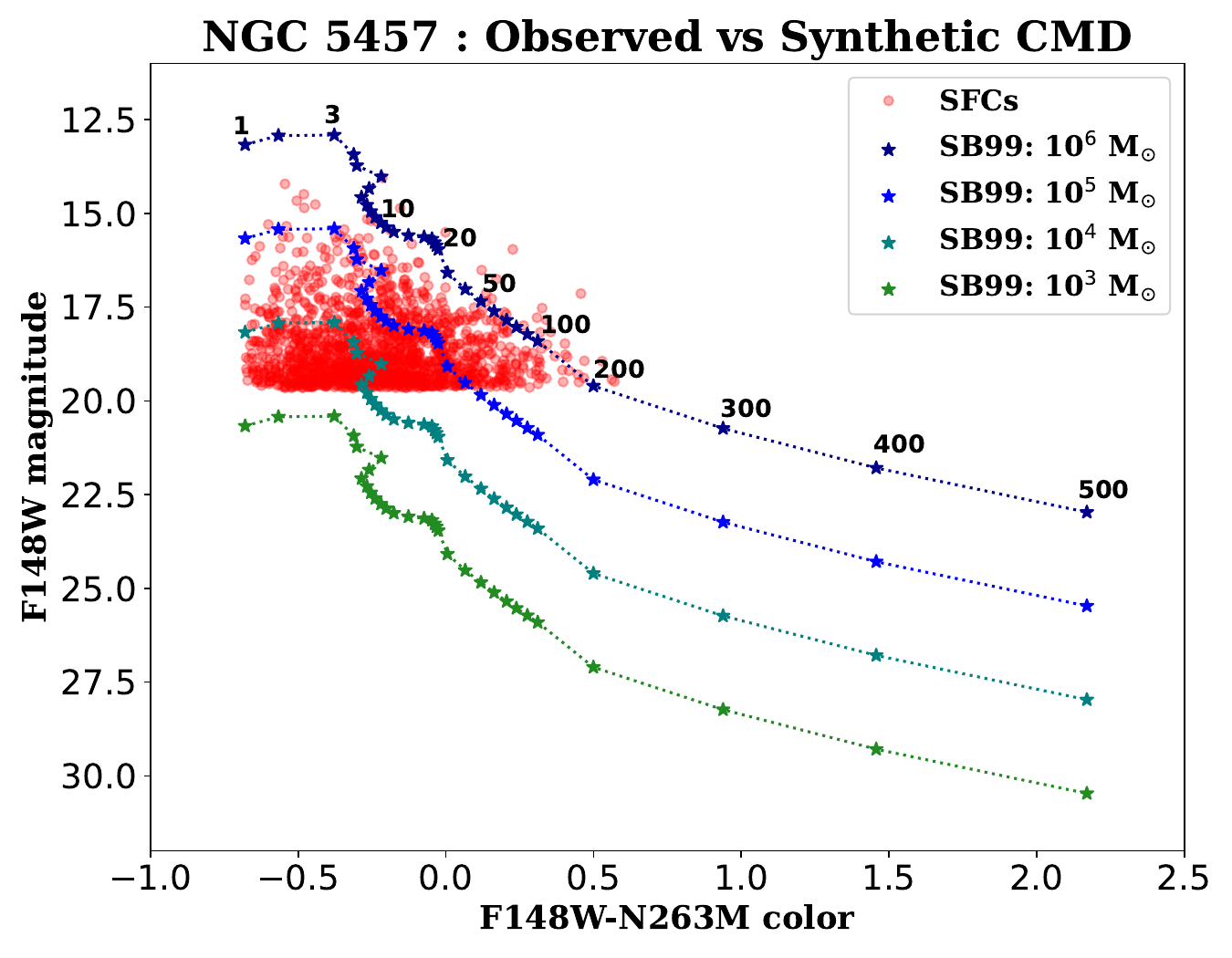}
		\label{fig:subfig2}
	    \end{subfigure}
     \caption{Synthetic CMDs from Starburst99 and the extinction-corrected SFCs identified in NGC 1566, NGC 5194 and NGC 5457. }
    \label{3_CMDs}
\end{figure}

\section{Toomre length estimates}
\label{appdx4}
We have tabulated the parameters used to estimate the Toomre length as well as the derived values of the Toomre length of the sample galaxies in Table \ref{Toom_table}. The molecular H$_2$ surface density ($\Sigma_\mathrm{g}$) values for NGC 1566, NGC 5194, NGC 5457, and NGC 7793 are taken from \citet{1995A&AS..114..147B}, \citet{1998ApJ...498..541K}, \citet{1998ApJ...498..541K}, and \citet{Israel_1995}, respectively. The flat rotation velocity ($v$) for NGC 1566, NGC 5194, NGC 5457, and NGC 7793 are taken from \citet{2019MNRAS.487.2797E}, \citet{deBlok_2008}, \citet{Guelin_1970}, and \citet{deBlok_2008}, respectively.\

\begin{table}
\caption{Summary of the parameters used in the Toomre length estimation and the resulting estimates for the sample galaxies.}
\centering
\begin{tabular}{ccccc}
\hline
Galaxy & $\Sigma_\mathrm{g}$  ($M_{\sun}$ $pc^{-2}$) & $v$ (km $s^{-1}$) & $r$ (kpc) & $l_{T}$ (pc)\\
\hline
NGC 1566 & 2.7 &  160 & 8.3 & 617\\
NGC 5194 & 24.0 & 220 & 5.5 & 1274\\
NGC 5457 & 1.7 &  180 & 8.5 & 322\\
NGC 7793 & 2.6 &  125 & 2.7 & 105\\\hline
\end{tabular}
\tablefoot{Molecular hydrogen density ($\Sigma_\mathrm{g}$) and flat rotation velocity ($v$) adopted directly from M21, median galactocentric radius of SFCs ($r$) and the estimated values of the Toomre length ($l_{T}$) of the sample galaxies.}
\label{Toom_table}
\end{table}

\section{Impact of local ultraviolet background on star-forming clump sizes}
\label{appdx5}
We found that the local UV background for our sample galaxies has an exponential profile, which approaches the sky background (bg) value at galactocentric radii much larger than the galaxy size. To roughly calculate the local UV background profile, we took circular annuli of 100 pixel width starting from the galactic centre and extending well beyond the full galaxy size; For each annulus, we subtracted the UV flux stored in the Astrodendro-identified clumps from the total UV flux to get the local UV background flux.\

\begin{figure}[ht!]
    \centering
    \includegraphics[width=0.45\textwidth]{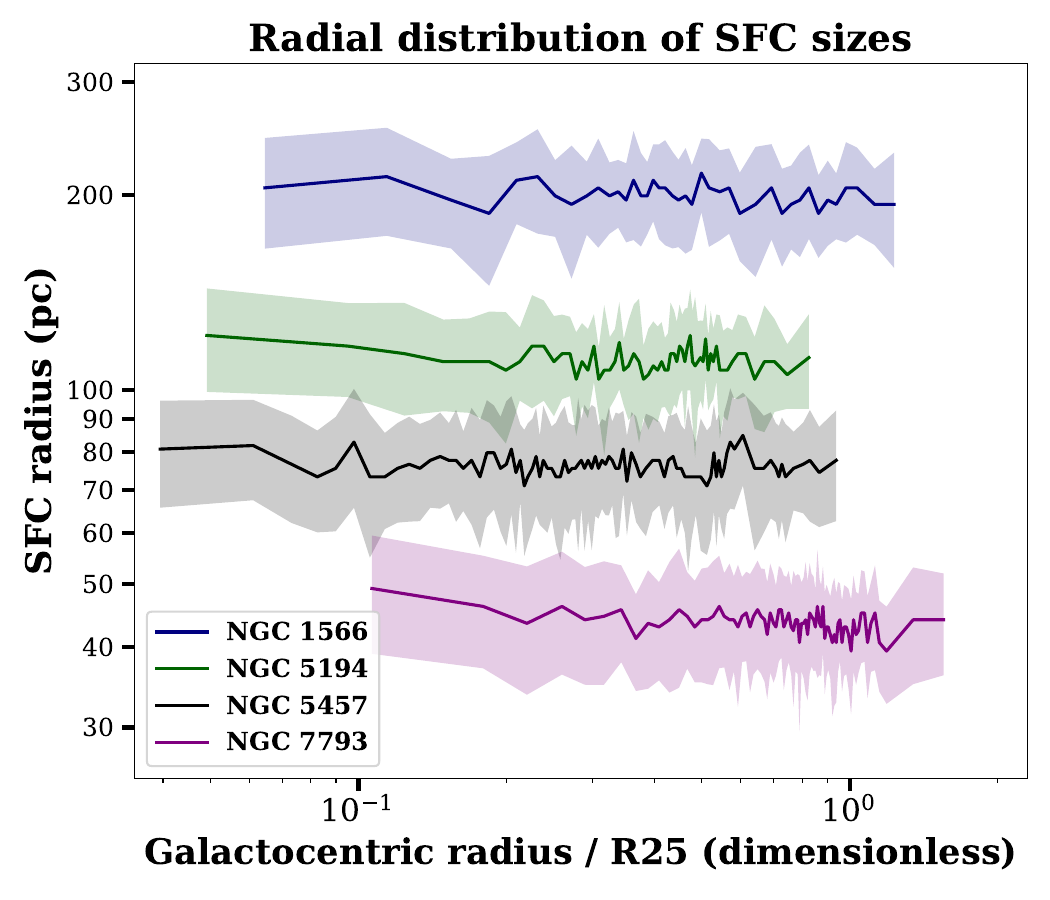}
    \caption{Radial size distribution of the SFCs identified by Astrodendro for our sample galaxies which shows that the sizes of the SFCs identified by Astrodendro roughly stay constant with galactoentric radius in our sample galaxies.}
    \label{radial size}
\end{figure}

We have used a fixed detection threshold (min\_value = 1 bg + 3 $\sigma$, based on bg and $\sigma$ values measured from the sky regions) in our Astrodendro analysis where the dominant contribution for setting the min\_value actually comes from 3 $\sigma$ since bg in FUV is usually much smaller than $\sigma$. SFC detection based on this fixed threshold may have some impact on the sizes of the SFCs in the regions where the detection threshold is much smaller than the local UV background. This condition can potentially be satisfied at smaller galactocentric radii where due to high local UV background, the SFCs can be clipped to larger sizes as compared to the outer disc. However, we found that for most of our sample galaxies, the min\_value chosen is higher than or comparable to the local UV background, even in the innermost regions of galaxies. This implies that the sizes of the detected SFCs should not be strongly affected by the local UV background. The radial distribution of the SFC sizes detected by Astrodendro (figure \ref{radial size}) clearly demonstrates this effect where the sizes of the SFCs identified by Astrodendro stays roughly constant with galactocentric radius, despite the radially varying local UV background. To create this figure, we plotted the SFC sizes versus their galactocentric radial position and created radial bins so that each bin contains approximately 100 SFCs. For each bin, we plotted the median SFC size and the standard error on the median size with respect to the median galactocentric radius. Thus, we can conclude that the effect of local UV background on the SFC sizes is minimal. \ 

\end{document}